\documentclass[aps, prx, letterpaper,twocolumn,superscriptaddress, showpacs, floatfix,longbibliography]{revtex4-1}
\usepackage{xcolor,bm,lineno,multirow,float,times}
\usepackage{physics}
\usepackage{rotating}
\usepackage{verbatim}
\usepackage{graphicx,tabularx}
\usepackage[multiple]{footmisc}
\usepackage[sort&compress]{natbib}
\usepackage[T1]{fontenc}
\usepackage{soul}

\newcommand{\blue}[1]{{#1}}

\def\ie{\textit{i.e. }}
\def\eg{\textit{e.g. }}

\usepackage[%
colorlinks=true,
 urlcolor=blue,
 linkcolor=blue,
 citecolor=blue
]{hyperref}

\begin{document}
	\title{Effective point-charge analysis of crystal fields -- application to rare-earth pyrochlores and tripod kagome magnets $R{_3}$Mg${_2}$Sb${_3}$O${_{14}}$}
	\author{Zhiling Dun}
	\affiliation{School of Physics, Georgia Institute of Technology, Atlanta, GA 30332, USA}
	\author{Xiaojian Bai}
	\affiliation{School of Physics, Georgia Institute of Technology, Atlanta, GA 30332, USA}
	\author{Matthew B. Stone}
	\affiliation{Neutron Scattering Division, Oak Ridge National Laboratory, Oak Ridge, TN 37831, USA}
	\author{Haidong Zhou}
	\affiliation{Department of Physics and Astronomy, University of Tennessee, Knoxville, TN 37996, USA}
	\author{Martin Mourigal}
	\affiliation{School of Physics, Georgia Institute of Technology, Atlanta, GA 30332, USA}
	\date{\today}
	\begin{abstract}
		An indispensable step to understand collective magnetic phenomena in rare-earth compounds is the determination of spatially-anisotropic single-ion properties resulting from spin-orbit coupling and crystal field (CF). The CF Hamiltonian has a discrete energy spectrum -- accessible to spectroscopic probes such as neutron scattering -- controlled by a number of independent parameters reflecting the point-symmetry of the magnetic sites. Determining these parameters in low-symmetry systems is often challenging. Here, we describe a general method to analyze CF excitation spectra using adjustable effective point-charges. We benchmark our method to existing neutron-scattering measurements on pyrochlore rare-earth oxides and obtain a universal point-charge model that describes a large family of related materials. We adapt this model to the newly discovered tripod Kagome magnets ($R_{3}$Mg$_2$Sb$_{3}$O$_{14}$, $R$ = Tb, Ho, Er, Yb) for which we report broadband inelastic neutron-scattering spectra. Analysis of these data using adjustable point-charges yields the CF wave-functions for each compound. From this, we calculate thermomagnetic properties that accurately reflect our measurements on powder samples, and predict the effective gyromagnetic tensor for pseudo-spin degrees of freedom -- a crucial step to understand the exotic collective properties of these kagome magnets at low temperature. We present further applications of our method to other tripod kagome materials and triangular rare-earth compounds $R$MgGaO$_4$ ($R$ =Yb, Tm). Overall, this study establishes a widely applicable methodology to predict CF and single-ion properties of rare-earth compounds based on interpretable and adjustable models of effective point-charges.
	\end{abstract}
	\maketitle

	\section{Introduction}
	In most magnetic insulators, electrons in the partially-filled atomic shells of transition-metal or rare-earth cations give rise to localized magnetic moments. In general, determining the individual properties of these magnetic moments is difficult because electrons' spins are coupled to their orbital angular momentum and to the surrounding environment of diamagnetic anions by the spin-orbit and Coulomb interactions. Furthermore, magnetic moments can interact with each other through exchange interactions, often via an intermediary ligand, or directly via dipole-dipole interactions. Nonetheless, in the low-energy limit, it is often possible to describe the magnetic dipole moment of these complex multi-electron systems as an \textit{effective} spin degree of freedom, with an anisotropic gyromagnetic tensor that stems from the spatial distribution of local magnetization, and an anisotropic bi-linear exchange Hamiltonian that describes the interactions between two nearby effective spins \cite{Vleck1932theory, jensen1991rare, abragam2012electron}. Anisotropic dipole moments and bond-dependent exchange interactions play a central role in forefront problems in quantum magnetism such as the realization of quantum spin-ice in pyrochlore systems \cite{Hermele_2004, Owen_2012, Gingras_2014}, Kitaev spin-liquids on the honeycomb lattice \cite{kitaev_2006, Singh_2012, banerjee_2016}, or triangular-lattice rare-earth antiferromagets such as YbMgGaO$_4$ \cite{Li_2015_PRL, shen2016evidence, Paddison_2017}. The first step to understand the magnetism of these frustrated magnets is to accurately capture their spin-space anisotropy, \ie properties of their gyromagnetic $g$-tensor and exchange tensors.
	
	For an isolated rare-earth ion, the spin-orbit interaction couples the total spin $S$ and orbital $L$ angular momenta of the unpaired electrons' manifold such that the total angular momentum $J\!=\!L\!+\!S$ is usually a good quantum number. When ions are embedded in a crystal, the 2$J$+1 level degeneracy is split by the electrostatic field produced by surrounding ligands, \eg the crystal field (CF). The theoretical framework of the CF theory was developed in 1952 by Stevens who first expressed the electrostatic potential of rare-earth ions as a linear combination of angular momentum operators, the Stevens' operators, from which the $g$-tensor can be directly obtained~\cite{stevens1952matrix}. In 1964, Hutchings demonstrated that CF energy levels can in principle be calculated from a point-charge ionic model of the ligand environment~\cite{hutchings1964point}. Later work showed that the number of CF parameters solely depends on the point-group symmetry of the ionic site, from 2 for cubic, 6 or 8 for hexagonal, to a maximum number of 26 for lower symmetry environments \cite{lea1962raising, walter1984treating}. For rare-earth systems, the energy scale of the CF generally varies between a few meV to hundreds of meV, depending on the nature of the ligands and their distance to the magnetic ions, which is at least one order of magnitude larger than exchange interactions. This separation of energy scale suggests that the collective multi-ion effects at low energies can be viewed as a perturbation to the high-energy single-ion physics; and all non-zero components of the exchange tensor can in principle be obtained from the CF wave-functions using a perturbation theory after considering the combination effects of space group symmetry, time reversal symmetry associated with Kramers/non-Kramers ions, and dipolar/multipolar nature of effective moments \cite{Onoda2011quantum, huang2014quantum,Rau2018frustration,Rau_2019}. In this sense, understanding spin-space anisotropy depends profoundly on the determination of the CF Hamiltonian and its parameters. 
	
	To date, \textit{ab-initio} calculations of CF effects for $f$-electron systems have not proven trustworthy. Therefore, the determination of the CF Hamiltonian in a real material relies heavily on interpretation and fitting of experimental data. Inelastic neutron scattering is one of the most advantageous experimental techniques to do so because it directly measures the CF eigenvalues (excitation energies) as well as the dipolar matrix elements between CF eigenfunctions (excitation intensities) \cite{moze1998crystal,furrer2009neutron}. Typically, the analysis consists in searching a high-dimensional space for CF parameters that best fit the experimental observables. In many cases, this process can be problematic: (i) the CF parameters are not directly associated with any measurable physical quantities, one usually does not know where to start within the high-dimensional parameter space; (ii) not all CF levels can be resolved experimentally due to low-intensity, mediocre resolution of neutron-scattering experiments at high energy-transfer or overlap with the phonon background; (iii) it is possible to encounter degenerate best-fit solutions yielding totally different CF wave-functions. The situation becomes especially challenging in low-symmetry materials where the number of experimental observables is considerably less than the number of CF parameters. A widely adopted strategy to resolve this problem is to start with a point-charge (PC) calculation by which the known positions of the surrounding ligands are used to estimate the CF parameters. Examples where such approach has been used include the pyrochlore Yb$_2$Ti$_2$O$_7$ \cite{Gaudet2015neutron}, and Nd-based tripod kagome compounds \cite{Scheie2018crystal}. However, as pointed out by Hutchings himself and as we will further demonstrate below, PC calculations based on a purely crystallographic model -- where point electric charges are placed at the crystal lattice sites -- have weaknesses because they neglect the finite extent of charges on the ions, covalent bonding with the ligands, and the complex effects of ``screening'' of the magnetic electrons by the outer electron shells of the magnetic ions \cite{hutchings1964point} . Therefore, CF calculations from the crystallographic PC model are usually not realistic. 
	
\begin{figure}[tbp!]
		\linespread{1}
		\par
		\begin{center}
			\includegraphics[width= \columnwidth ]{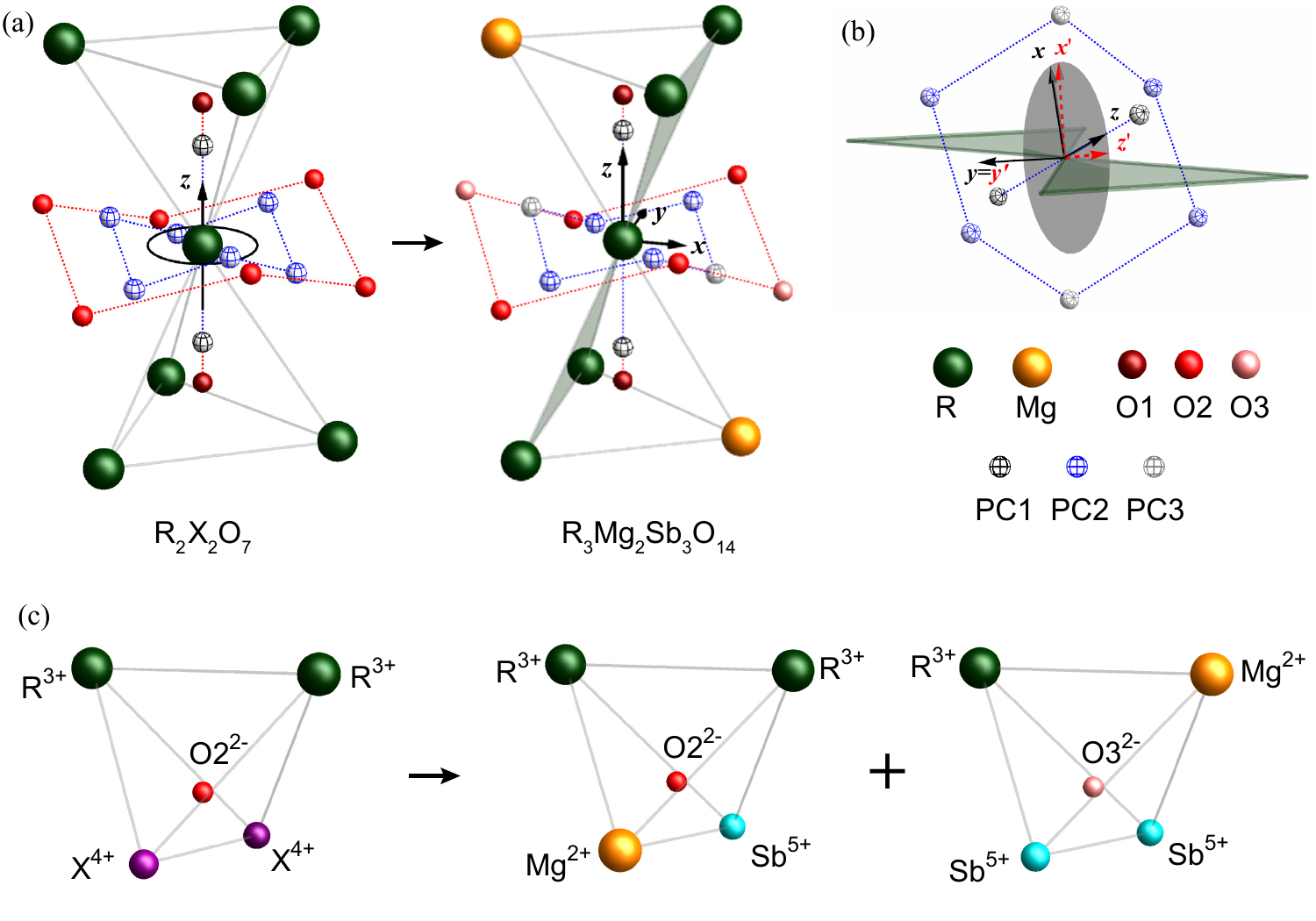}
		\end{center}
		\par
		\caption{\label{Fig:Structure} (a) Partial crystal structure of rare-earth pyrochlore $R{_2}$X${_2}$O${_{7}}$ and tripod kagome magnets $R{_3}$Mg${_2}$Sb${_3}$O${_{14}}$ where $R$ stands for rare earth ion. Solid spheres represent atomic ions, and nonequivalent oxygen ions are differentiated by colors. Empty spheres represent effective point-charges (PC) as described in section III. For $R{_2}$X${_2}$O${_{7}}$, a continuous XY spin anisotropy is illustrated as a black circle. (b) A side view of the eight PC (empty spheres) in a tripod kagome structure with respect to the kagome plane (green area). The local $xyz$ coordination is chosen such that the $y$-axis is the two-fold rotation axis $C_2$ of the $C_{2h}$ point group, and the $z$-axis is along the $R$-O1 bond (or $R$-PC1) direction. A finite rotation about the $C_2$-axis is required to obtain a new $x'y'z'$ frame corresponding to the principal axes for which the $g$-tensor is diagonal. The rotational plane is highlighted by a grey disk. (c) Differences in local environment for the six oxygen atoms in the puckered ring (joined by red dashed lines in (a)) between the pyrochlore and tripod kagome structure. The differences call for a modification of the effective-PC model, as described in Sec.~\ref{sec:tripod}. }
	\end{figure}

	In this manuscript, we adopt a new approach to solve this problem. Instead of fitting the CF excitations using Steven's operators, we employ a direct calculation and fitting algorithm based on an effective-PC model, which relies on point electric-charges located on the rare-earth-ion to ligand segment and carrying a reduced charge. This model overcomes the weakness of PC calculations in a semi-empirical way, and has recently been successfully used in the community of single-molecule magnets \cite{baldovi2012modeling, SIMPRE}. The advantage of our approach is a physically meaningful parameterization and a constrained parameter space. This proves to be extremely valuable for low-symmetry systems, such as the tripod kagome magnets \cite{Dun_2016, Sanders_2016, Scheie2016effective, Paddison_2016, Dun_2017, dun2020quantum} presented in this study [Fig.~\ref{Fig:Structure}] for which the number of required parameters is reduced from 15 to 9. 
	
	The structure of the paper is as follows. First, in Sec.~\ref{sec:methods}, we describe the experimental methods as well as and the theoretical principles involved. Second, in Sec.~\ref{sec:pyrochlore}, we introduce the concept of effective-PC Model and provide a benchmark to the the existing inelastic neutron scattering measurements of pyrochlore rare-earth oxides. Then, in Sec.~\ref{sec:tripod}, we proceed with the main experimental results of this work, the CF excitations of the newly discovered tripod kagome magnets ($R_{3}$Mg$_2$Sb$_{3}$O$_{14}$, $R$ = Tb, Ho, Er, Yb) \cite{Dun_2017}. We modify the effective-PC model and perform fit to the data, from which susceptibility, isothermal magnetization, $g$-tensor, and principal axes are obtained and compared to the experimental observations.  Finally, we discuss some other applications of the effective-PC model, including understanding the pressure effects on the transverse field, as well as making predictions for other tripod kagome materials and triangular compounds $R$MgGaO$_4$ ($R$= Yb, Tm).  \blue{All the data along with the Python3 code used for the point charge fit is available online at Github \cite{PointChargeCF_2021} for the community to test and benchmark.} In short, although our work does not provide new physical insights into the CF theory, it describes a general methodology to modify the PC model to more accurately analyze and predict CF phenomena in real materials.    
	
	\begin{table*}[tbp]  
		\setlength{\tabcolsep}{0.6em} 
		\begin{tabular}{c|cccc|cccc|cccc}
			\hline\hline 
			\centering
			&\multicolumn{4}{c|}{$R_2X_2$O$_7$} & \multicolumn{4}{c|}{$R_3$Mg$_2$Sb$_3$O$_{14}$} & \multicolumn{4}{c}{$R$MgGaO$_4$}\\
			\hline
			No.	&  $r$(\AA) & $\theta(^\circ)$ & $\phi(^\circ)$ & $q$($e$) & $r$(\AA) & $\theta(^\circ)$ & $\phi(^\circ)$ & $q$($e$) & $r$(\AA) &  $\theta(^\circ)$ & $\phi(^\circ)$ & $q$($e$) \\ 
			\hline
			1	&  $r_1$ & 0              & 0    &  $q_1$ &  $r_1$ & 0              & 0            &  $q_1$  & $r_1$ & $\theta_1$     & 0   &  $q_1$  \\
			2	&  $r_1$ & 180            & 0    &  $q_1$ &  $r_1$ & 0              & 0            &  $q_1$  & $r_1$ & $\theta_1$     & 120 &  $q_1$ \\
			3	&  $r_2$ & $\theta_2$     &  60  &  $q_2$ &  $r_2$ & $\theta_2$     & $\phi_2$       &  $q_2$  & $r_1$ & $\theta_1$     & 240 &  $q_1$ \\
			4	&  $r_2$ & 180-$\theta_2$ & 120  &  $q_2$ &  $r_2$ & 180-$\theta_2$ & 180-$\phi_2$ &  $q_2$  & $r_1$ & 180-$\theta_1$ & 60  &  $q_1$ \\
			5	&  $r_2$ & $\theta_2$     & 300  &  $q_2$ &  $r_2$ & $\theta_2$     & 360-$\phi_2$ &  $q_2$  & $r_1$ & 180-$\theta_1$ & 180 &  $q_1$\\
			6	&  $r_2$ & 180-$\theta_2$ & 240  &  $q_2$ &  $r_2$ & 180-$\theta_2$ & 180+$\phi_2$ &  $q_2$  & $r_1$ & 180-$\theta_1$ & 300 &  $q_1$\\
			7	&  $r_2$ & $\theta_2$     & 180  &  $q_2$ &  $r_3$ & $\theta_3$     & 180          &  $q_3$ \\
			8	&  $r_2$ & 180-$\theta_2$ &   0  &  $q_2$ &  $r_3$ & 180-$\theta_3$ &   0          &  $q_3$ \\
			\hline
			PC parameters &\multicolumn{4}{c|}{$r_1, r_2, \theta_2, q_1, q_2$ } & \multicolumn{4}{c|}{$r_1, r_2, r_3, \theta_2, \theta_3, \phi_2, q_1, q_2, q_3$ }  & \multicolumn{4}{c}{$r_1, \theta_1, q_1$ } \\
			\hline
			\multirow{ 2}{*}{CF parameters} &\multicolumn{4}{c|}{\multirow{ 2}{*}{$A^0_2, A^0_4, A^3_4, A^0_6, A^3_6, A^6_6$} } & \multicolumn{4}{c|}{$A^0_2, A^1_2, A^2_2, A^0_4,A^1_4, A^2_4,A^3_4, A^4_4$}  & \multicolumn{4}{c}{\multirow{ 2}{*}{$A^0_2, A^0_4, A^3_4, A^0_6, A^3_6, A^6_6$} } \\
			&\multicolumn{4}{c|}{} & \multicolumn{4}{c|}{$ A^0_6, A^1_6, A^2_6, A^3_6, A^4_6, A^5_6, A^6_6$}  & \multicolumn{4}{c}{} \\	
			\hline
			Crystallographic PC model &\multicolumn{12}{c}{$r_i = r_i^c $,\quad $\theta_i = \theta_i^c$ ,\quad $\phi_i = \phi_i^c$, \quad $q_i = 2e$ }  \\
			\hline
			\multirow{ 2}{*}{Effective PC model} &\multicolumn{12}{c}{$r_i = f*r_1^c $, \quad $\theta_i = \theta_i^c$ ,\quad $\phi_i = \phi_i^c$ }  \\
			&\multicolumn{4}{c|}{$q_1 = 0.5e$, $q_2 = 0.333e$   } & \multicolumn{4}{c|}{$q_1 = 0.5e$,\quad $q_2 = 0.3e$,\quad $q_3 = 0.15e$  }  & \multicolumn{4}{c}{$q_1 = 0.5e$ } \\
			\hline\hline  	
		\end{tabular}
		\caption{\label{table:PC} Parametrization of Point-Charge (PC) models written in spherical coordinates for the point-group symmetry of rare-earth pyrochlores $R_2X_2$O$_7$ ($D_{\rm 3d}$), tripod kagome magnet $R_3$Mg$_2$Sb$_3$O$_{14}$ ($C_{\rm 2h}$), and triangular-lattice rare-earth antiferromagnets $R$MgGaO$_4$ ($D_{\rm 3d}$). The last case describes a rare-earth ion in the center of an octahedron with 3-fold symmetry which also applies to the site-disordered $R$ ions in $R_2X_2$O$_7$ and $R_3$Mg$_2$Sb$_3$O$_{14}$. The amount of charge is measured in the unit of electron charge ($e$). The variables $r_i^c$, $\theta_i^c$, and $\phi_i^c$ denote the crystallographic parameters associated with the rare-earth ion to oxygen-ligand parameters in real materials. For the effective PC model, $f$ denotes a reduction factor associated with the shortest $R-$O bond distance, and we find $f \approx$ 0.72 is a proper choice for most materials in this study. }
	\end{table*}
	
	\section{Methods}
	\label{sec:methods}
	
	\subsection{Sample Synthesis}
	All the tripod kagome compounds $R{_3}$Mg${_2}$Sb${_3}$O${_{14}}$ ($R$ = Tb, Ho, Er, Yb) were synthesized by a sol-gel technique using rare-earth oxides (Tb$_4$O$_{11}$, Er$_2$O$_3$, Ho$_2$O$_3$, Yb$_2$O$_3$, 99.9\%), MgO (99.99\%), Sb$_2$O$_3$ (99.9\%), nitric acid (ACS grade), tartaric acid (C$_4$H$_6$O$_6$), and citric acid (C$_6$H$_5$O$_7$) as starting materials. For each compound, stoichiometric ratios of $R$(NO$_3$)$_3$, Mg(NO$_3$)$_3$ (prepared by dissolving rare earth oxides and MgO in hot diluted nitric acid solution), and antimony tartarate (prepared by dissolving Sb$_2$O$_3$in hot tartaric acid solution) were first mixed in a beaker. Citric  acid with a metal-to-citric  molar ratio of 1:2 was then added to the solution followed by a subsequent heating on a hot plate at 120$^\circ$C overnight to remove excessive water. The obtained gel-like solution was slowly heated to 200$^\circ$C in a box furnace to decompose the nitrate, and was pyrolyzed at 600$^\circ$C for 12 hours in air. The obtained powder was ground up, pressed into a pellet and re-heated at 1300 to 1350$^\circ$C until a well reacted crystallized powder was obtained. It is noteworthy that powder samples of tripod kagome compounds synthesized by the conventional solid state reactions are usually accompanied by 2-3\% magnetic impurities (mainly a robust $R{_3}$SbO${_7}$ phase \cite{fennell2001structural}), which had been a considerable complication for interpreting the thermodynamic properties of the system in previous studies~\cite{Paddison_2016}. A big advantage of the sol-gel synthesis is its high efficiency in getting impurity-free samples. A comparison between measurements on two Ho${_3}$Mg${_2}$Sb${_3}$O${_{14}}$ samples synthesized by the two methods can be found in Ref.~\onlinecite{dun2020quantum}.
	
	\subsection{Experimental Measurements}
	Inelastic neutron-scattering measurements were performed on the fine-resolution Fermi chopper spectrometer (SEQUOIA) \cite{Granroth_2010} at the Spallation Neutron Source (SNS), Oak Ridge National Laboratory (ORNL), USA. For each of the tripod kagome compounds, a powder sample with a typical mass $\approx$5\,g was loaded in an aluminum container (an aluminum annular cylinder was used for Er${_3}$Mg${_2}$Sb${_3}$O${_{14}}$ to minimize absorption), and was cooled down to 5\,K using a closed-cycle refrigerator.  Data were collected with incident neutron energies $E_\textrm{inc.}\!=\!240$, $120$, and $30$\,meV (yielding an elastic energy-resolution of 5.5, 1.9, and 0.5\,meV, respectively) at temperatures $T\!=\!300$\,K, $100$\,K, $50$\,K, and $5$\,K. The same measurements were repeated for an empty aluminum sample holder and used for background subtraction. Data reduction was performed using the Mantid \cite{Arbold2014} to yield the neutron scattering intensity $I(Q,\omega)$ as a function of momentum-transfer $Q$ and energy-transfer $\hbar\omega$. Data were further processed with the DAVE program \cite{Azuah_2009}. The phonon contribution to the scattering intensity was subtract in two different ways. For the $E_\textrm{inc.}\!=\!30$~meV datasets, we take advantage of the absence of low-energy CF excitations in Yb${_3}$Mg${_2}$Sb${_3}$O${_{14}}$ below $E\!\approx\!50$~meV and use its spectra as phonon background for the other three compounds. For the higher $E_\textrm{inc.}$ datasets, phonon background was modeled and subsequently subtracted by assuming a $Q^2$ intensity dependence. 
	
	Magnetic susceptibility and isothermal magnetization were measured using a Quantum Design Physical Properties Measurement System (PPMS). Magnetic susceptibility, $\chi(T)= M(T)/H$, was measured while cooling the sample from $T\!=\!400$~\,K to $1.8$\,K in an external field $\mu_0H\!=\!0.1$~T.  Isothermal magnetization, $M(H)$, were measured between 0 $\le\mu_0H\le14$~T at selected temperatures between $T=1.8$\,K and $40$\,K. The measured were was corrected for the diamagnetic background of the sample holder which is crucial for the values of 1/$\chi(T)$ at high temperatures. Diamagnetic contribution from the sample is much smaller, therefore is not corrected in this study.
	
	\subsection{Point-Charge Calculations}
The electrostatic potential experienced by $f$-electrons can be expended in a series of polynomials of order sixth or lower, such that the CF Hamiltonian has the form,
\begin{equation}
\mathcal{H}_{\mathrm{CF}} = \sum_{n,m} \left[ A^{m}_{n}  \theta_n \right] O^m_n= \sum_{n,m}B^{m}_{n} O^m_n, \label{eq:Steven}
\end{equation} 
where $O^{m}_{n}\,(m \leq n)$ are the Stevens' operators \cite{stevens1952matrix, elliott1953theory}, $A^{m}_{n}$ and $B^{m}_{n}$ are the CF parameters. Here, $ \theta_{n}$ represent reduced matrix elements that have been previously calculated for each $R^{3+}$ ion in Ref \cite{stevens1952matrix} . 
	
	Following the method outlined by Hutchings~\cite{hutchings1964point}, the CF levels can be calculated on the basis of a simple PC model where the electrostatic potential is approximated by a sum over the Coulomb potentials from $N$ surrounding PCs at positions $\mathbf{R}_i$ with charges $q_i$ (in the unit of electron charge, $e$), \ie $V(r,\theta,\phi) = \sum_{i} q_i/\abs{\mathbf{R}_i- \mathbf{r}}$. When expressing the potential in tesseral harmonics $Z_{nm}$, the PC Hamiltonian becomes:
	\begin{eqnarray} \label{eq:Hutching}  
	\mathcal{H}_{\mathrm{CF}}  &=&  -\sum_{i} q_i e \sum_{n,m} r^n \gamma^{nm}_i Z_{nm}(r, \theta,\phi),   \\ \nonumber 
	\gamma^{nm}_i  &=&  \frac{4\pi}{2n+1} Z_{nm}(\mathbf{R}_i)/R^{n+1}_i.
	\end{eqnarray}  
	By connecting the expression of $Z_{nm}$ to $O^{m}_{n}$ \cite{hutchings1964point}, the $A^{m}_{n}$ parameters are determined by the following expression:
	\begin{equation}
	A^{m}_{n} = -\sum_{i} C^{m}_{n} \langle r^n \rangle \gamma^{nm}_i q_i , \label{eq:Amn}
	\end{equation}   
	where $C^{m}_{n}$  are the prefactors of the spherical harmonics,  $\gamma^{nm}_i$ is given by Eq. \ref{eq:Hutching},  \blue{and $\langle r^n \rangle$ is the expectation value of the  $4f$-electron radial wave function that have been tabulated for each $R^{3+}$ ion \cite{Freeman1979, edvardsson1998role}. Here, we stick to the convention used in Ref. \cite{SIMPRE, furrer2009neutron} where the factor $\langle r^n \rangle$ is absorbed in $A^{m}_{n}$. Note this factor is  excluded in some other references, \eg Ref. \cite{bertin2012crystal, Scheie2018crystal, gaudet2018effect}.}
	
	From here, it is clear that within the PC approximation, Eq.~\ref{eq:Hutching} and Eq.~\ref{eq:Amn} allow one to construct the CF Hamiltonian (Eq. \ref{eq:Steven}) from the coordination and charge of surrounding ligands. In that context, the CF parameters can be interpreted as a summation over the tesseral harmonics coefficients for the $N$ surrounding PCs of the ligands. Because the number of CF parameters required to describe $\mathcal{H}_{\mathrm{CF}}$ solely depends on the point-group symmetry of the rare-earth site, as tabulated by Walter \cite{walter1984treating}, it is not directly related to the number of independent PC variables associated with the surrounding ligands. As a consequence, the PC model shows clear advantages over the conventional Stevens' operators approach: (i) unlike the CF parameters ($A^{m}_{n}$ or $B^{m}_{n}$), the PC variables ($\mathbf{R}_i$, $q_i$) are more physically meaningful; (ii) thus, the number of free parameters can be greatly reduced using explicit physical or chemical constraints; (iii) through PC calculations, one can easily track the changes of the CF properties induced by modifications to the crystallographic structure. These advantages are particularly important in the case of low symmetry systems, such as the tripod kagome compounds discussed further below.
	
	\begin{figure*}[tbp!]
		\linespread{1}
		\par 
		\begin{center}
			\includegraphics[width= 5.5 in ]{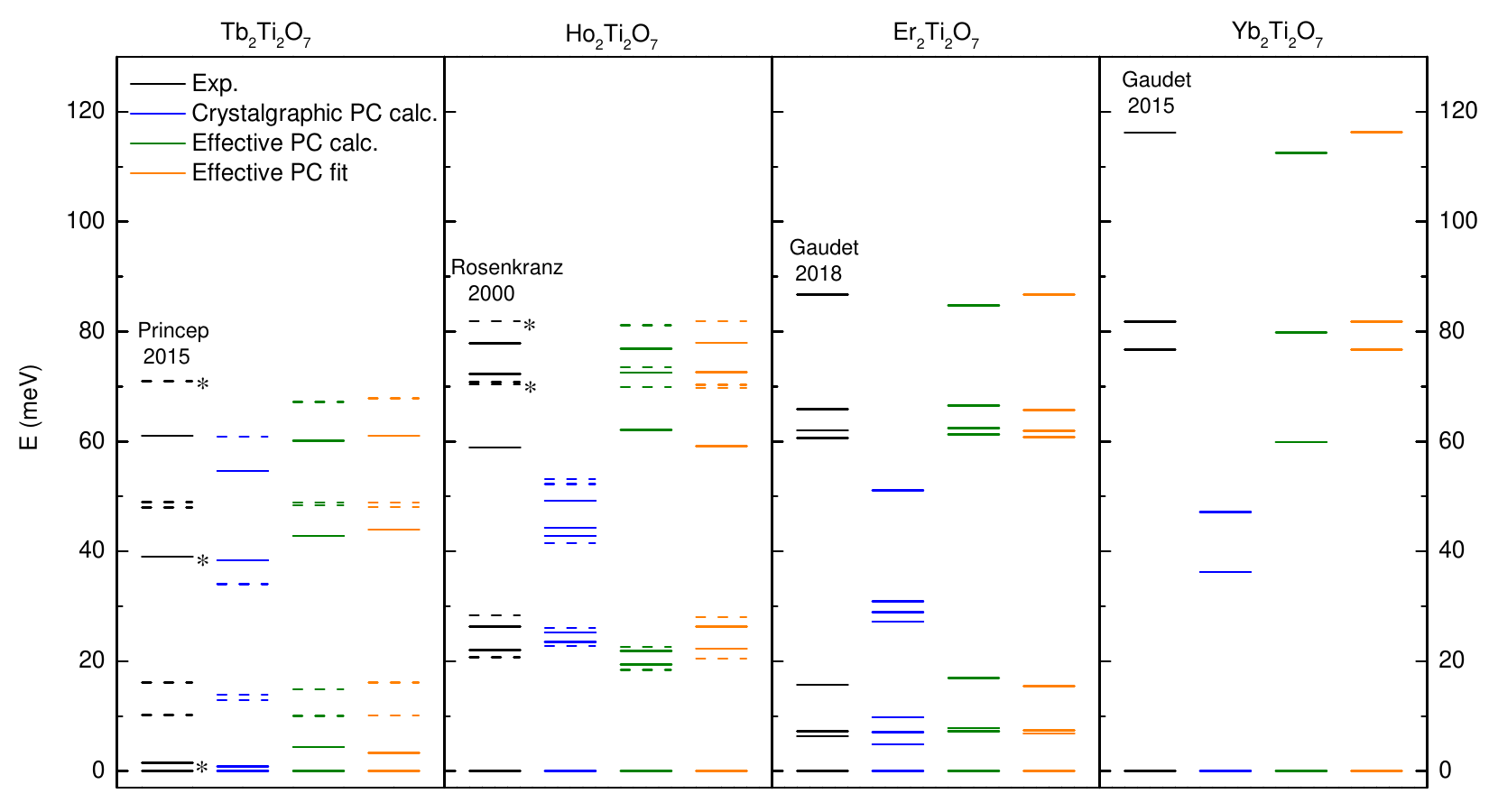}
		\end{center}
		\par
		\caption{\label{Fig:Pyrochlorelevels} Comparison between the CF levels measured by inelastic neutron-scattering \cite{princep2015crystal, Rosenkranz_2000, gaudet2018effect, Gaudet2015neutron} and the calculated/fitted CF levels from different PC model for rare-earth pyrochlore titanates $R{_2}$Ti${_2}$O${_{7}}$ ($R$ = Tb, Ho, Er, Yb). See Sec.~\ref{sec:pyrochlore} and Tab.~\ref{table:PC} for the definitions of our PC models. Dashed and solid lines represent singlet and doublet CF levels, respectively. The CF levels that are not directly observed experimentally are marked by stars, and are not included in the PC fit. }
	\end{figure*}
	
	\subsection{Point-charge Fit to CF excitations}
	For a rare-earth ion with total angular momentum $J$, the eigenstates of Eq.\,\ref{eq:Steven} are 2$J$+1 levels with eigenvalues $E_n$ and eigenvectors expressed in the total angular momentum basis as $\ket{\Gamma_n} = \sum_{j=-J}^{J} C_{n,j}\ket{J,J_z = j}$. 
	
	Inelastic neutron-scattering probes the magnetic-dipole-active transitions between these levels. Within the dipole approximation, the powder averaged neutron-scattering intensity is: 
	\begin{equation}  \label{eq:Sqw}
	\begin{split}
	I(Q,\omega)= C F^{2}(Q) \sum_{n, m} &\frac{\sum_{\alpha=x,y,z}\left|\left\langle \Gamma_n\left|J_{\alpha}\right| \Gamma_{m}\right\rangle\right|^{2} \mathrm{e}^{-E_{n}/k_\mathrm{B}T}}{\sum_{j} \mathrm{e}^{-E_{j}/k_\mathrm{B}T}} \\ &\times\delta(\hbar\omega + E_n - E_m), 
	\end{split}
	\end{equation}
	where $C$ is a constant,  $F^{2}(Q)$ is the squared magnetic form factor, $k_B$ is the Boltzmann constant, $E_{n}$ and $E_{m}$ are the  eigenvalues of the CF Hamiltonian, and $\hbar\omega$ is the neutron energy transfer \cite{furrer2009neutron}. Each measurement was performed at fixed temperature $T$ and incident neutron energy $E_\textrm{inc.}$ . The measured $I(Q,\omega)$ was integrated within a certain $Q$ range (see individual plots below) and subsequently normalized to its maximum intensity to obtain the $\tilde{I}(\omega)$ that we plot further below.
	
	To analyze $\tilde{I}(\omega)$, we start from a set of PC parameters \{$\mathbf{R}_i$, $q_i$\}, use Eqns.~\ref{eq:Hutching} and \ref{eq:Amn} to construct the CF Hamiltonian, and diagonalize Eq.~\ref{eq:Steven} to obtain the eigenvalues $E_n$ and eigenfunctions $\ket{\Gamma_n}$. Several available program packages are capable of doing such calculations, including SIMPRE \cite{SIMPRE} (used in this study), McPhase \cite{rotter2004using}, and PyCrystalField \cite{scheie2021pycrystalfield}. With obtained $A_n^m$ parameters, we  diagonalize Eq. \ref{eq:Steven} to get eigenstates and eigenenergies \cite{PointChargeCF_2021},  then use Eq. \ref{eq:Sqw} to calculate the neutron scattering intensity by replacing the Dirac $\delta$-function with a Voigt function:
\begin{equation} \label{eq:Voigt}
	\begin{split}
& V(\omega ; \sigma_G, \gamma_L)  \\
& \equiv   \int_{-\infty}^{\infty}   G\left(x ; \sigma_G\right) L\left(\hbar\omega + E_n - E_m-x ; \gamma_L\right) d x,
\end{split}
\end{equation}	
 \blue{
 where $G$ is a Gaussian function to account for the  energy-resolution ($\sigma_G$)  of neutron scattering spectrometer  which is energy-dependent for SEQUOIA, and $L$ is a Lorentzian function with $\gamma_L$ representing the intrinsic broadening (or finite lifetime) of CF excitations. 
 }
  By varying the $\mathbf{R}_i$ and $q_i$ variables of our PC model along with $\gamma_L$, a least-squares fit is performed to minimize the difference between calculated and observed CF spectra. The agreement is measured by a self-defined weighted profile factor that includes all data-sets measured with different $E_\textrm{inc.}$ and $T$ :
	\begin{equation} \label{eq:R}
	R_{\mathrm{wp}} =  \frac{1}{N} \sqrt{\sum_{i} \left(\frac{I_i^{\text{calc}}-I_i^{\text{obs}}}{\sigma_i^\text{obs}}\right)^2},
	\end{equation} 
	where $I_i^{\text{calc}}$, $I_i^{\text{obs}}$, $\sigma_i^{\text{obs}}$ represent calculated intensity, observed intensity, and measurement error, respectively, for the $N$ data points.  
	 Due to the high dimensionality of the parameter space, the Nelder-Mead method \cite{nelder1965simplex} was adopted to search for local minimum in the parameter space whereas the  choice of initial $\mathbf{R}_i$ and $q_i$ will be discussed below in Secs.~\ref{sec:pyrochlore} and ~\ref{sec:tripod}.
	
\subsection{Susceptibility \&  Magnetization }
	Static magnetic properties in an external magnetic field $\mathbf{H}$ can also be calculated from the single-ion CF Hamiltonian,
	\begin{equation} \label{eq:MagH}
	\mathcal{H} =  \mathcal{H}_{\mathrm{CF}} - \mu_{B} g_{J}  \mathbf{H} \cdot \mathbf{J}. 
	\end{equation} 
	With the eigenstate ($E_{n}$) and eigenfunction ($\ket{\Gamma_n}$) of the CF Hamiltonian available, the three components ($\alpha=x,y,z$) of the magnetization $\mathbf{M}^\text{CF}(\mathbf{H}, T)$ in a Cartesian coordinate system are given by
	\begin{equation} \label{eq:Mxyz}
	M^\text{CF}_\alpha(\mathbf{H}, T) = g_J\sum_{n} e^{-\frac{E_n}{k_{\rm B} T}} \bra{n}J_\alpha\ket{n} /\sum_{n} e^{-\frac{-E_n}{k_{\rm B} T}}, 
	\end{equation}
	from which the DC magnetic susceptibility tensor can be calculated numerically following
	\begin{equation} \label{eq:Chixyz}
	\chi_{\alpha\beta}  =\frac{\partial M_{\alpha}}{\partial H_{\beta}} . 
	\end{equation}
	Within a linear response regime,  $\chi_{\alpha\beta}$ remains a constant for small $\mathbf{H}$, such that the powder-averaged magnetic susceptibility in a suitable choice of $x, y, z$ axes is:
	\begin{equation} \label{eq:Chipowder}
	\chi_{\text {powder}}^\text{CF}  = \frac{1}{3}(\chi_{xx}+\chi_{yy}+\chi_{zz}). 
	\end{equation}
	Outside the linear response regime, the powder-averaged isothermal magnetization $M_{\text{powder}}(H, T)$  for a polycrystalline sample can be calculated numerically assuming randomly oriented structural domains, 
	\begin{equation}  \label{eq:MH_powder}
	M_{\text {powder}}^\text{CF}(H, T) = \frac{1}{4\pi}\int_{0}^{2\pi}\!\int_{0}^{\pi} (\mathbf{M}^\text{CF}(H,T)\cdot \hat{\mathbf{H}})\sin\theta d\theta d\phi.
	\end{equation}
	
	The above susceptibility and magnetization calculations only contain CF contributions and neglect exchange and dipolar interactions between magnetic ions. In the temperature regime for which two-ion interactions are non-negligible, the corrections to the susceptibility and magnetization can be largely accounted by a Weiss molecular field \cite{weiss1907hypothese}. That is, we assume that each magnetic ion experiences a local field proportional to the magnetization in the paramagnetic region, 
	\begin{equation}  \label{eq:molecule-field}
	\mathbf{H}_\textrm{Weiss} = \lambda \mathbf{M}^{\text{CF}}(\mathbf{H}_\text{Weiss}+\mathbf{H}, T),
	\end{equation}
	where $\lambda$ is a constant that reflects the average magnetic couplings between ions. Given a value of $\lambda$, Eq. \ref{eq:molecule-field} can be solved self-consistently to find the local molecular field ($\mathbf{H}_\textrm{Weiss}$) given the temperature ($T$) and the external field ($\mathbf{H}$). By replacing $\mathbf{M}^{\text{CF}} (\mathbf{H}, T)$ by the corrected magnetization $\mathbf{M}^{\text{CF}+\lambda} (\mathbf{H}, T) = \mathbf{M}^{\text{CF}}(\mathbf{H}_\text{Weiss}+\mathbf{H}, T)$ in Eqns.~\ref{eq:Chixyz}-\ref{eq:MH_powder}, we obtain the corrected powder-averaged magnetic susceptibility $\chi_{\text {powder}}^{\text{CF}+\lambda}$, and isothermal magnetization $\mathbf{M}_\text{powder}^{\text{CF}+\lambda} (\mathbf{H}, T)$. In the limit of high temperature and small field, the correction to the susceptibility takes a simple form of the Weiss law, 
	\begin{equation}  \label{eq:Curie-Weiss}
	1/{\chi_{\text {powder}}^{\text{CF}+\lambda}} = 1/{\chi_{\text {powder}}^{\text{CF}}} +\lambda.
	\end{equation}

\subsection{principal axes \& effective $g$-tensor}
In rare-earth oxides, the CF energy-scale is usually much larger than two-ion exchange and dipolar interactions. When the spectrum of the single-ion Hamiltonian yields a group of ground-state eigenstates that are well separated from excited levels, the interactions between magnetic moments expressed in the total angular-momentum basis can be projected into the ground-state subspace at low temperature because only these states are thermally populated. In this context, it is desirable to use the concept of ``effective spin",$\mathbf{S}$ (or sometimes referred as pseudo spin), which is a fictitious angular momentum such that 2$S$+1 is set equal to the degeneracy of the single-ion ground-state. In the case where the CF ground-state is a doublet, we construct effective spin-1/2 operators from the CF doublet wave-functions in the total angular momentum basis ($\ket{\pm}$), following
\begin{eqnarray}
	S^{\alpha} = \frac{1}{2}\left(\begin{array}{cc}
	{\left\langle+\left|J_\alpha\right|+\right\rangle} & {\left\langle-\left|J_\alpha\right|+\right\rangle} \\ {\left\langle+\left|J_\alpha\right|-\right\rangle} & {\left\langle-\left|J_\alpha\right|-\right\rangle}\end{array}\right), \alpha=x,y,z. \label{eq:pseudospin}
\end{eqnarray} 
The effective spin is thus connected to the the Pauli matrices ($\sigma_\alpha$) by  an anisotropic $g$-tensor,
\begin{eqnarray} \label{eq:g-factor}
	\left(\begin{array}{l}{S_{x}} \\ {S_{y}} \\ {S_{z}}\end{array}\right)
	= \frac{\hbar}{2}\left(\begin{array}{lll}{g_{xx}} & {g_{xy}} & {g_{xz}} \\ {g_{yx}} & {g_{yy}} & {g_{yz}} \\ {g_{zx}} & {g_{yz}} & {g_{zz}}\end{array}\right)\left(\begin{array}{l}{\sigma_{x}} \\ {\sigma_{y}} \\ {\sigma_{z}}\end{array}\right),
\end{eqnarray} 
from which we obtain the mapping of the Zeeman splitting under an external magnetic field ($\mathbf{H}$) from the total angular-momentum basis to the pseudo-spin basis \blue{where the  $g$-tensor is contained: }
\begin{eqnarray}
	\mathcal{H}_{Zee} = -\mu_{B} g_{J} \mathbf{H}\cdot\mathbf{J}  \; \longmapsto  -\mu_{B}  \mathbf{H}\cdot\mathbf{S}. \label{eq:g-factormap}
\end{eqnarray}
	
In an arbitrarily chosen coordinate system, $\bm{g}$ is an 3$\times$3 tensor. Our aim is to find a principal coordinate system such that $\bm{g}$ is diagonal. In axial symmetry, we can choose the $z$-axis as the local symmetry axis so that all the off-diagonal terms vanish, and the $g$-tensor can be rewritten as $g_{xx}$ = $g_{yy}$ = $g_\perp,\;g_{zz} = g_{\parallel} $. This is exactly the case for the rare-earth pyrochlores and triangular-lattice compounds discussed in this study. If we choose the local 3-fold local axis (the $R$-O1 bond direction) as the $z$-axis, $\bm{g}$ is automatically diagonal despite the choices of $xy$-axes, meaning a continuous rotational symmetry is preserved for the single-ion magnetism [Fig .\ref{Fig:Structure}(a)]. In contrast, one expects $g_{xx} \neq g_{yy} \neq g_{zz}$ in lower-symmetry systems for which it remains a technical challenge to determine $\bm{g}$ as well as its principal axes. In the tripod kagome structure, one principal axis is the local $C_2$ rotation axis (labeled as $y$ in Fig.~\ref{Fig:Structure}). Unlike the pyrochlores, the other two principal axes, $x$ and $z$, are undetermined and can in principle lie anywhere within the plane perpendicular to $y$ (illustrated as a grey plane in Fig. \ref{Fig:Structure}(b)). Presumably, the $z$ axis is likely to be the shortest $R$-O1 bond direction due to its structural similarity to the pyrochlore structure. However, PC calculations show that the directions of principal axes strongly depends on the details of surrounding ligands. Generally, it requires a finite rotation about the $C_2$ axis to make $\bm{g}$ diagonal, as illustrated in Fig. \ref{Fig:Structure}(b).            
	
	We show here that the principal axes and $\bm{g}$ can be determined by a two-step rotation \cite{chibotaru2008unique}. Starting from the CF wave-functions $\ket{\pm}$, the first step is a pseudo-spin rotation of Eq.~\ref{eq:g-factor} with rotation matrix $A$,
	\begin{eqnarray} \label{eq:rotationA}
	\bm{S} = \frac{\hbar}{2} (\bm{g} A^{-1}) (A \bm{\sigma}) = \frac{\hbar}{2} \bm{g'}\bm{\sigma'}. 
	\end{eqnarray}
	This rotation has nothing to do with the rotation in real space which is rather a rotation of pseudo-spin to make $\bm{g'}$ symmetric. The second step is a co-rotation of real space and pseudo-spin space through a rotation matrix $B$,
	\begin{eqnarray} \label{eq:rotationB}
	B\bm{S} = \frac{\hbar}{2} (B\bm{g'} B^{-1}) (B \bm{\sigma'}). 
	\end{eqnarray}
	For the tripod kagome compounds, if we start with $\ket{\pm}$ from the PC calculation using the $xyz$ coordination shown in Fig. \ref{Fig:Structure}, $B$ will be the rotation about the $C_2$ axis which ultimately determines the  principal axes $x'y'z'$ as well as the diagonalized effective $g$-tensor.

	\section{Rare-earth pyrochlores: a benchmark}
	\label{sec:pyrochlore}
	
	Before applying our PC calculations and analysis to new material systems, it is desirable and necessary to validate our method on a well-studied family of compounds. The rare-earth pyrochlores serve as a perfect testing ground; first, because their structure is closely related to that of our target systems, the tripod kagome magnets; second, because their CF Hamiltonian is relatively simple with only six Stevens' operators due to the presence of a 3-fold symmetry axis; finally and most importantly, because their CF excitations have been intensively investigated by neutron scattering over the last two decades, providing a complete and reliable set of data \cite{Rosenkranz_2000, Mirebeau2007magnetic, princep2013crystal, Zhang2014neutron, princep2015crystal, Jaubert_2015, ruminy2016crystal, gaudet2018effect, gaudet2018magnetoelastically}. 
	
	\subsection{Crystallographic point-charge model}
	Rare-earth pyrochlore oxides possess a general chemical formula of $R_2$X$_2$O$_7$ ($X$ = Ti, Sn, Ge, Pt, Zr, etc.) with space group $Fd\bar{3}m$, and point group $D_{3d}$ at the rare-earth ion site. Each $R^{3+}$ is surrounded by eight oxygen atoms with two short $R$-O1 bonds lying along the local three-fold axis and six long $R$-O2 bonds forming a puckered ring [Fig. \ref{Fig:Structure}]. It requires three independent parameters to fully describe the coordination of the eight oxygens. In spherical coordinates where $z$ is chosen as the 3-fold axis, these are $r_1$, $r_2$, $\theta_2$, representing the crystallographic distance of the $R$-O1 bond, $R$-O2 bond, and the  O1-$R$-O2 angle, respectively.  Modeling the oxygen ligands by eight PCs requires two additional variables, $q_1$ and $q_2$, that describe the amount of charge associated with O1 and O2. The minimal PC model thus contains five parameters for 8 PCs (see Tab.~\ref{table:PC} for details).
	
	The most intuitive choice of PC parameters is to adopt the crystallographic ligand-charge positions, \ie $r_1 = r_1^c$, $r_2 = r_2 ^c$, $\theta_2 = \theta_2^c$, and net charges of isolated $O^{2-}$ ions, \ie $q_1 = q_2 = 2e$. This model is the electrostatic point-charge model first considered by Bethe in 1929 \cite{bethe1929termaufspaltung}, which we call the \textit{Crystallographic PC model}. In Fig. \ref{Fig:Pyrochlorelevels}, we compare the CF levels calculated from this model with the experimentally measured CF levels for the rare-earth titanates $R_2$Ti$_2$O$_7$. The crystallographic PC model generally underestimates the overall CF energy scales for all systems; and more dramatically, it predicts a Ising-like $g$-tensor ($g_\parallel \gg g_{\perp}$) for Er$_2$Ti$_2$O$_7$ while the actual spin anisotropy in the real compound is known to be XY-like ($g_\parallel < g_{\perp}$).  We notice that the second-order CF term predicted by the model is too large while the fourth and sixth order terms are considerably smaller in comparison to the experimental values. This is not a surprise given the known weaknesses of the PC model~\cite{zolnierek1984crystal} and calls for a modification to the crystallographic PC model.
	
	\begin{figure}[!tbp]
		\linespread{1}
		\par 
		\begin{center}
			\includegraphics[width= \columnwidth ]{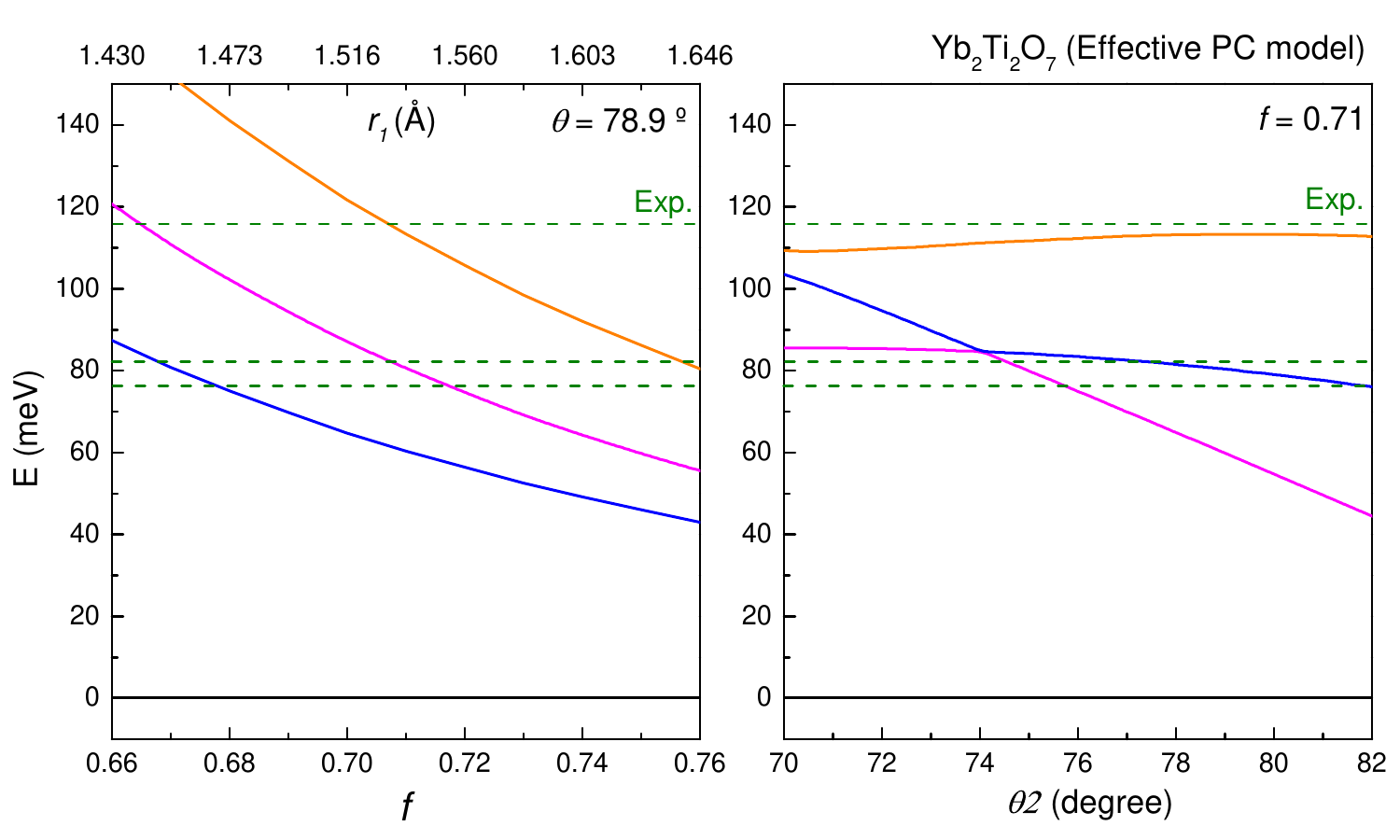}
		\end{center}
		\par
		\caption{\label{Fig:Dependence} Dependence of the CF energies as a function of rare-earth to ligand distances and angles in an effective PC model for Yb$_2$Ti$_2$O$_7$ (Tab.~\ref{table:PC}).  Experimentally observed levels from Ref. \cite{Gaudet2015neutron} are marked by green dashed lines.}
	\end{figure}
	\subsection{Effective point-charge model}
	Many efforts have been undertaken over the years to correlate the CF parameters derived form the PC model with experimental observations, including introducing a shielding parameter~\cite{sternheimer1968shielding}, adding dipolar and quadrupolar electric potential corrections \cite{hutchings1963investigation}, and taking into account the electro-negativity \cite{zolnierek1984crystal} and wave-functions overlap \cite{porcher1999relationship} of the metal and ligand ions. These improvements rely on the admittance of following effects: the finite extent of charges on the ions, the contribution from the rest of the crystalline net, and the covalency between the metal and ligand wave-functions. Interestingly, all these modified PC models make the following adjustments to the PC parameters: first, the effective charges carry a considerably reduced charge compared to bare ones; second, rather than being strictly localized at the crystallographic ligand centers, the effective PCs are placed somewhere in the middle of the metal-ligand bond. The difference between the various models is their attempts to relate the reduction in charge and distance to semi-empirical physical quantities, \eg electro-negativity \cite{zolnierek1984crystal}, and wave-function overlap \cite{porcher1999relationship, SIMPRE}. However, we find that none of these semi-empirical approaches work well for the pyrochlore compounds discussed below. Thus, in this study, we adopt the general concept of \textit{effective PC model} by which each ligand carries an effective PC with adjustable charge amount and  distance to the metal ion.
	
	We start from the local geometry of the pyrochlore structure to build up the effective PC model. As shown in Fig.~\ref{Fig:Structure}(c), both O1 and O2 are in a tetrahedral environment, \blue{where the spatial distribution of charge density is expected to contain the same symmetry as  the $sp^3$ hybridization of a CH$_4$ molecule}. However, O1 is at the center of a regular tetrahedron formed by four $R^{3+}$ ions while O2 is inside a irregular tetrahedron formed by two $R^{3+}$ and two $X^{4+}$ ions.  Since the covalency of O$^{2-}$ with $R^{3+}$ is different from that of $X^{4+}$, it becomes necessary to distinguish the effective PCs associated with O1 and O2. If the 2$e$ charge amount of O$^{2-}$ is distributed on a tetrahedron based on the amount of positive charge on the surronding ions, then on average O1 contributes 0.5$e$ to each $R^{3+}$ while O2 contributes 0.333$e$ to each $R^{3+}$ and 0.667$e$ to each $X^{4+}$, \ie $q_1$ = 0.5$e$, $q_2$ = 0.333$e$ [Tab.~\ref{table:PC}]. This simple counting argument ensures that the total negative charges due to ligands is balanced with the positive charge of metallic ions. 
	
	Under these assumptions, we find that a single effective distance parameter is sufficient to describe the CF excitations reasonably well. This yields a reduction factor $f$ associated with the shortest crystallographic metal–ligand distance used, \ie $r_1 = r_2 = f*r_1^c$. According to the electro-negativity argument~\cite{zolnierek1984crystal}, $f = \epsilon_R/(\epsilon_R+\epsilon_O)\approx 0.75$, where $\epsilon_R$ and $\epsilon_O$ are the Pauling electro-negativity of the rare-earth  and oxygen ions, respectively. By fixing $q_1$ = 0.5$e$, $q_2$ = 0.333$e$ and varying $f$, we find that the best agreement with experiment observations is achieved for $f \approx 0.72$, for which the calculated CF levels are in close agreement with the measured levels (green lines in Fig.~\ref{Fig:Pyrochlorelevels}).  Perhaps more importantly, this effective PC model predicts the correct spin anisotropy of the CF ground-state wave-functions for all four compounds (see Tab.~\ref{table:Pyrochlores}).
	Since the effective PC model is still defined by the local crystallography, the agreement with experiments is encouraging and shows that a universal and physically meaningful PC model can achieved for the series rare-earth pyrochlore titanate compounds. 
	
	\begin{figure}[tbp]
		\linespread{1}
		\par 
		\begin{center}
			\includegraphics[width=  \columnwidth ]{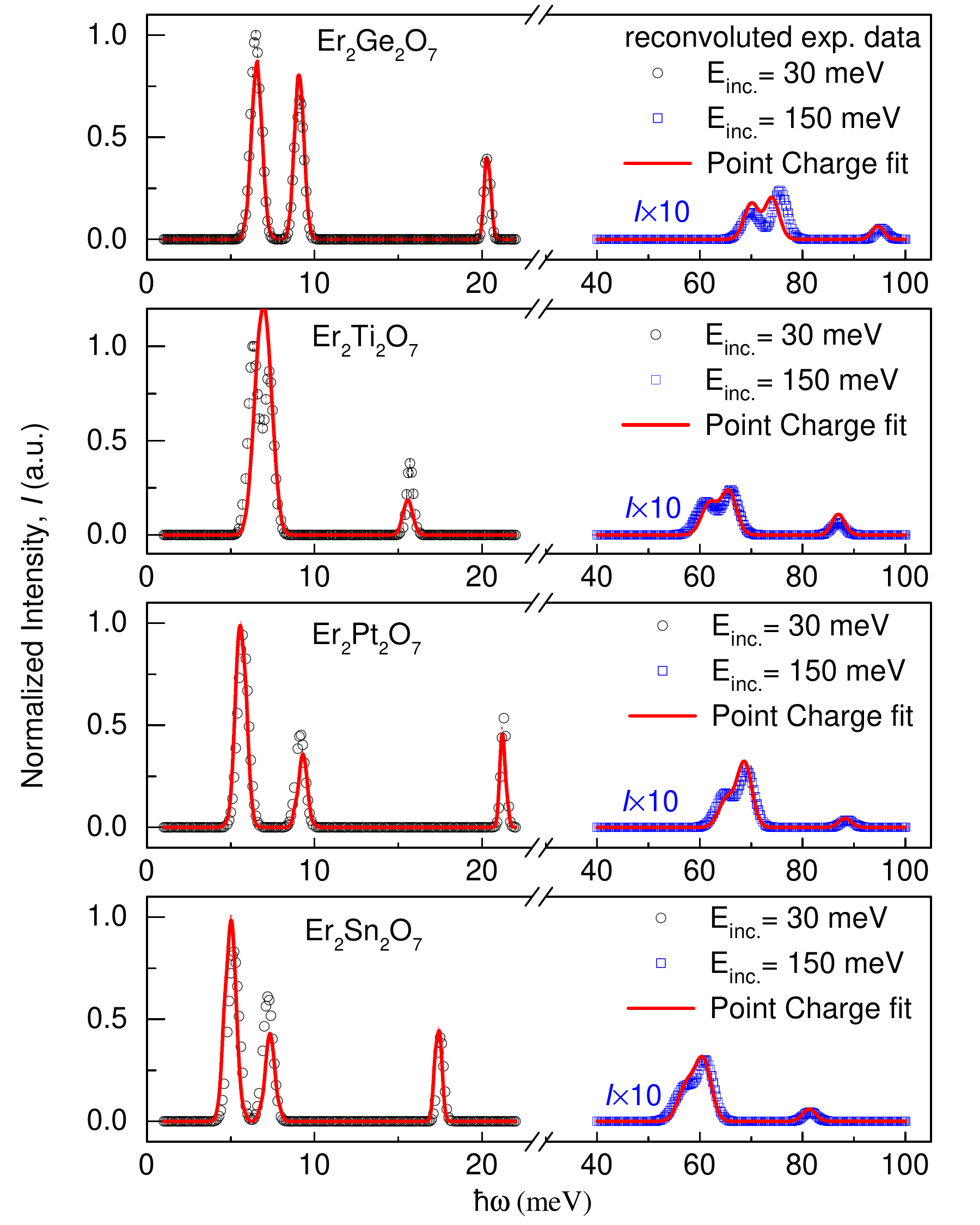}
		\end{center}
		\par
		\caption{\label{Fig:Er-pyrochlores} Best point charge fit results (red lines) to the normalized neutron CF excitation spectra of Er$_2X_2$O$_7$ ($X$ = Ge, Ti, Pt, Sn). Experimental data (open symbols) are adapted from Gaudet \textit{et al.}~\cite{gaudet2018effect} where each experimentally observed CF excitation is re-convoluted to a Voigt function with a $\Omega_G$ = 0 meV, and $\Omega_L$ chosen to be the instrument resolution of SEQUOIA  according to $E_\textrm{inc.}$ = 30 meV and 150 meV, respectively.  }
	\end{figure}
	
	\begin{table*}[tbp]  
		\setlength{\tabcolsep}{0.45em} 
		\centering
		\caption{\label{table:Pyrochlores} \blue{ Calculated, fitted parameters and $g$-tensors from the effective PC model for rare-earth pyrochlores. These numbers are compared to those from conventional Stevens Operator fits from Ref. \cite{princep2015crystal, Rosenkranz_2000, Gaudet2015neutron, gaudet2018effect}. For $R_2$Ti$_2$O$_7$ ($R$ = Tb, Ho, Er, Yb), the Effective PC parameters are defined in Table.\ref{table:PC} with $f$ = 0.72, and the calculated CF levels are plotted as green lines in Fig. \ref{Fig:Pyrochlorelevels}. For each compound of the Er$_2X_2$O$_7$ ($X$ = Ti, Ge, Sn, Pt) family, a fit to the re-convoluted experimental neutron scattering data was performed with the five PC parameters as variables, as shown in Fig. \ref{Fig:Er-pyrochlores}.  }}
		\begin{tabular}{c|ccccc|c|cccccc|c|c}
			\hline\hline 
			\multirow{2}{*}{Compound} &	\multicolumn{5}{c|}{ PC parameters} & Ratio & \multicolumn{6}{|c|}{CF parameters (meV) } &$g$-tensor &  \multirow{2}{*}{Method}\\
			 & $r_1$(\AA) & $r_2$(\AA) & $\theta_2 (^\circ)$ & $q_1 (e)$ & $q_2 (e)$ & $\frac{q_1}{r_1}$/$\frac{q_2}{r_2}$ & $A_2^0$ & $A_4^0$ & $A_4^3$ & $A_6^0$ & $A_6^3$ & $A_6^6$ & $g_\parallel$ \;  $g_\perp$  &  \\
			\hline 
			Tb$_2$Ti$_2$O$_7$  & 1.638 & 1.638 & 79.5 & 0.5 & 0.333 & 1.5 & 19.8 & 39.5 & 318.6 & 5.9 & -91.0 & 98.7 & 11.4\; 0.0 &  PC Calc.(this work)\\
			                                                                                    &  &  &  &  & & & 27.6 & 46.3 & 378.6 & 7.1 & -146.0 & 114.0 &  11.2\; 0.0 & Stevens Op. fit \cite{princep2015crystal} $^a$\\
			\hline 			
			Ho$_2$Ti$_2$O$_7$ & 1.594 & 1.594 & 79.4 & 0.5 & 0.333 & 1.5 & 19.2 & 38.6 & 313.0 & 5.6 & -87.5 & 94.5 & 19.5 \; 0.0 & PC Calc. (this work)\\
			                                                                                    &  &  &  &  & & & 34.1 & 34.35 & 247.6 & 5.4 & -80.0 & 96.5.0 &  19.6\; 0.0 & Stevens Op. fit \cite{Rosenkranz_2000} $^b$\\		
			\hline 
			Yb$_2$Ti$_2$O$_7$ & 1.540 & 1.540 & 78.9 & 0.5 & 0.333 & 1.5 & 15.2 & 37.5 & 321.9 & 5.6 & -85.8 & 89.4 & 2.7 \; 3.7 & PC Calc.(this work)\\
			                                                                                 &  &  &  &  & & & 35.8 &  35.5 & -181.9 &   7.4 &   250.0  & 33.8 &  1.9\; 3.6 & Stevens Op. fit \cite{Gaudet2015neutron} $^b$\\		
			 \hline 
			 Er$_2$Ti$_2$O$_7$& 1.571 & 1.571 & 79.1 & 0.5 & 0.333 & 1.5 & 19.8 & 38.5 & 322.9 & 5.8 & -89.0 & 94.0 & 0.4 \; 7.1 & PC Calc.(this work)\\
			                                   & 1.627 & 1.656 & 78.8 & 0.646 & 0.461 & 1.43 & 20.4 & 41.3 & 351.5 & 6.1 & -85.9 & 89.1 & 3.0 \; 6.5 & PC fit (this work)\\			 
			                                                                                 &  &  &  &  & & & 26.7 &  43.8 & 369.1 &   6.0 &   -82.6 & 104.0 &  3.9\; 6.4 & Stevens Op. fit \cite{gaudet2018effect} $^c$\\
		    \hline                                                                            
			Er$_2$Ge$_2$O$_7$ & 1.568 & 1.545 & 82.1 & 0.567 & 0.350 & 1.60 & 19.2 & 47.6 & 274.7 & 5.0 & -83.8 & 116.7 & 3.8 \; 6.4 & PC fit (this work)\\
			                                                                                 &  &  &  &  & & & 27.9 &  47.4 & 360.0 &    5.9 &  -92.0 & 128.1 &  3.9\; 6.3 & Stevens Op. fit \cite{gaudet2018effect} $^c$\\		
			\hline 
			Er$_2$Sn$_2$O$_7$ & 1.608 & 1.558 & 80.2 & 0.535 & 0.302 & 1.72 & 27.5 & 37.8 & 276.9 & 4.7 & -79.9 & 92.4 & 1.2  \; 7.2 & PC fit (this work) \\
			                                                                                &  &  &  &  & & & 36.1 &  37.2 & 348.2    & 5.3 & -82.6 & 100.7 &  0.1\; 7.6 & Stevens Op. fit \cite{gaudet2018effect} $^c$\\			
			\hline 
			Er$_2$Pt$_2$O$_7$ & 1.603 & 1.514 & 79.6 & 0.540 & 0.285 & 1.80 & 27.5 & 40.2 & 290.1 & 4.8 & -91.1 & 108.2 & 1.0  \; 7.8 & PC fit (this work) \\
																							&  &  &  &  & & & 34.9 &  40.8 & 343.0    & 5.3 & -92.2 & 118.5 &  0.3\; 7.7 & Stevens Op. fit \cite{gaudet2018effect} $^c$\\				\hline\hline  
			\multicolumn{15}{l}{\footnotesize{$^a$ Wybourne normalised paramters ($B_m^n, n>m$) in Ref. \cite{princep2015crystal} have been converted to Stevens Normalisation ($B_n^m, n>m$).}} \\
			\multicolumn{15}{l}{\footnotesize{$^b$  CF Parameters $B_n^m$ have been converted to $A_n^m$ following  Eq.~\ref{eq:Steven} }}\\
			\multicolumn{15}{l}{\footnotesize{$^c$  The number of $A_n^m$ in Ref. \cite{gaudet2018effect} has been multiplied by a factor of  $\left\langle r^n \right\rangle$ due to different definitions.  }}
		\end{tabular}
	\end{table*}
	
	
	\subsection{Point-charge fit}
	Starting from the effective PC model discussed above, we embark on the investigation of the effects of ligand-metal distances ($r_1$, $r_2$) and angle ($\theta_2$) on the CF levels.  An example is presented in Fig.~\ref{Fig:Dependence} for Yb$_2$Ti$_2$O$_7$, which shows that the overall CF energy scale is determined by the ligand-metal distance whereas the detailed splitting of CF levels is strongly affected by the angular distribution of the PCs. As expected, once we allow the five PC parameters ($r_1$, $r_2$, $\theta_2$, $q_1$, $q_2$) to vary slightly around the values defined by the effective PC model, excellent fits to the CF levels can be achieved (orange lines in Fig.~\ref{Fig:Pyrochlorelevels}). 
	
	As the number of fitted PC parameters is one less than the number of CF parameters (see Tab.~\ref{table:PC}), a natural question arises whether the fitted PC model reflects the nature of CF wave-functions in the real compounds. Previous studies on Er-based pyrochlores Er$_2X_2$O$_7$ ($X$ = Ge, Ti, Pt, Sn) \cite{gaudet2018effect} offer a perfect testing ground for these questions because Er$^{3+}$, with $J$ = 15/2, exhibits 8 CF doublets, which along with the scattering intensities provide solid constrains to unambiguously determine the CF Hamiltonian. More importantly, previous studies have shown that the CF ground-state wave-functions are delicately tuned by chemical pressure from the non-magnetic $X$ site, giving rise to distinct effective $g$-tensors \cite{gaudet2018effect}. As listed in Table \ref{table:Pyrochlores}, while all compounds exhibit XY anisotropy, the $g$-tensor of Er$_2$Ti$_2$O$_7$ and Er$_2$Ge$_2$O$_7$ are closer to the Heisenberg point while Er$_2$Pt$_2$O$_7$ and Er$_2$Sn$_2$O$_7$ are much more anisotropic with $g_\parallel \gg g_{\perp} $. To see whether the anisotropy of the effective $g$-tensor can be derived accurately from our PC fit, we reconstruct the CF excitations at $T$ = 5\,K for each Er$^{3+}$ compound based on the energy levels and scattering intensities from Ref. \cite{gaudet2018effect}. Our perform PC fit to the reconstructed data following the methods of Sec.~\ref{sec:methods}, and obtain an satisfactory agreement [Fig. \ref{Fig:Er-pyrochlores}]. The CF parameters as well as the $g$-tensors calculated from the fit results are listed in Tab.~\ref{table:Pyrochlores} and are very close to the values obtained using conventional Stevens' operator approaches (Eq.~\ref{eq:Steven}). Moreover, a rough estimation of the Coulomb potential between oxygens O1 and O2 ($\frac{q_1}{r_1}/\frac{q_2}{r_2}$) provides a possible explanation for the distinct $g$-tensors in the four compounds: the Coulomb potential associated with the puckered-ring of O2 has been greatly reduced in Er$_2$Pt$_2$O$_7$/Er$_2$Sn$_2$O$_7$ compared to that of Er$_2$Ti$_2$O$_7$/Er$_2$Ge$_2$O$_7$, which makes the $g$-tensor more anisotropic. In the limit of $\frac{q_2}{r_2} \to 0$, our PC calculations show that the ground state CF wave-functions will take the simplest form $\ket{\pm} = \ket{J_z = \pm1/2}$ so that  $g_\parallel = 9.6$ and  $g_{\perp} = 1.2$. Moreover, we find that whereas the crystallographic Er-O1 distance is always smaller than that of Er-O2 for all compounds, a good fit to experimental data requires $r_1/r_2 >1$ except for $X$ = Ti. Furthermore, the larger the $X$ atomic number, the larger the fitted $r_1/r_2$ ratio. A possible explanation is that the covalency betwen oxygen and $X$ is very different between $X^{4+}$ ions with empty $d$-orbitals (\eg Ti$^{4+}$) and ions with filled $d$-orbitals (\eg Ge$^{4+}$, Sn$^{4+}$, Pt$^{4+}$). It suggests that although the Coulomb potential of these non-magnetic ions is not explicitly contributing to the CF in the PC model, it is nonetheless reflected in the parameters of our effective PC model and thus can tune the single-ion properties.
	
	In short, by examining the existing experimental data for two families of rare-earth pyrochlore oxides, our results provide a solid benchmark of the effective PC model, showing that: (i) by combining the effects of local crystallography and covalency, it is possible to derive a physically meaningful and universal effective PC model that faithfully reproduces the CF spectra for a series of compounds; (ii) because the effective PC model reduces the number of free parameters compared to the traditional approach, a least-squares fit to experimental data allows to faithfully predict the spatially-anisotropic single-ion properties of a given system.
	
	\section{Tripod Kagome magnets}
	\label{sec:tripod}
	In this section, we turn to the tripod kagome magnets. We first consider the site symmetry as well as the local environment of the tripod kagome structure and arrive at a modified effective PC model. Next, we present the experimental results from the inelastic neutron scattering measurements on $R_{3}$Mg$_2$Sb$_{3}$O$_{14}$ ($R$ = Tb, Ho, Er, Yb), and perform PC fits to their CF excitation spectra.  We check the validity of our PC fit by comparing calculated magnetic susceptibility and isothermal magnetization to experiments. Finally, we calculate the $g$-tensor of the CF ground-state for each compound, and discuss the implications for the collective multi-ion physics. 
	
	\begin{figure*}[tbp]
		\linespread{1}
		\par
		\begin{center}
			\includegraphics[width= 6.5in]{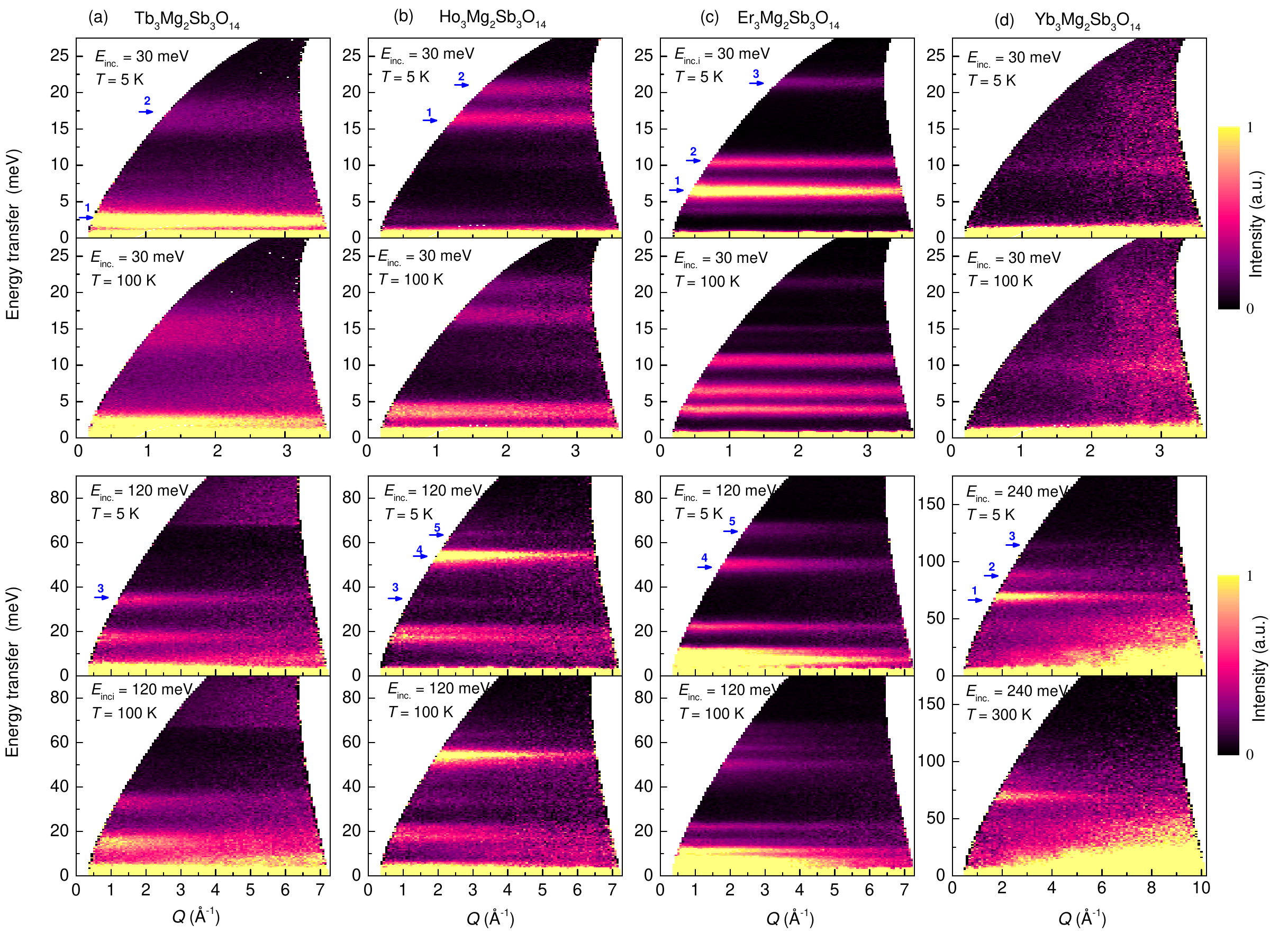}
		\end{center}
		\par
		\caption{\label{Fig:INS}  Momentum- and energy-dependence of the inelastic neutron scattering intensity $I({\bf Q},\omega)$ for $R_{3}$Mg$_2$Sb$_{3}$O$_{14}$ ($R$ = Tb, Ho, Er, Yb) measured  with different neutron incident energies at different temperatures. The top two rows show the $E_\textrm{inc.}$ = 30 meV datasets at $T=$5\,K and 100\,K, respectively. The bottom two rows shows the $E_\textrm{inc.}$ = 120 meV datasets except for Yb$_{3}$Mg$_2$Sb$_{3}$O$_{14}$ for which $E_\textrm{inc.}$ = 240 meV was used. Distinct groups of excitations from the CF ground-state are marked by blue arrows.}
	\end{figure*}
	
	\subsection{From structure to effective PC model}
	The crystal structure of the tripod kagome compounds $R_{3}$Mg$_2$Sb$_{3}$O$_{14}$ can be viewed as a variant of the pyrochlore which contains kagome planes of magnetic rare-earth ions separated by nonmagnetic Mg$^{2+}$ triangular layers [Fig.~\ref{Fig:Structure}(a)]. The space-group has changed from cubic F$d\bar{3}$m in the pyrochlores to trigonal group R$\bar{3}$m in the tripod systems. Importantly, although each $R^{3+}$ ion is still surrounded by eight oxygen atoms, the site symmetry is reduced from $D_{3d}$ to $C_{2h}$ \cite{Dun_2017}. This can be seen directly from the local structure, which instead of having one three-fold and three two-fold rotational axes, only preserves axial symmetry with a two-fold rotational $C_2$ axis that lies in the kagome plane [Fig.~\ref{Fig:Structure}(b)]. As a consequence, whereas it requires only 6 Stevens' operators to describe the CF Hamiltonian of the pyrochlores, it calls for 15 CF parameters for the tripod kagome magnets \cite{walter1984treating}. If we choose the $y$-axis as the $C_2$ axis, these are $A_n^m$ (or $B_n^m$) with $n = 2,4,6$ and $m = 0,1,...,n$. For many of the compounds in the tripod-kagome family, determining the parameters of the CF Hamiltonian directly from neutron-scattering spectra is impossible because the experimental observables are considerably fewer than the fitting parameters. For example, only three excitation levels and two intensity ratios can be extracted from the CF excitations for Yb$_{3}$Mg$_2$Sb$_{3}$O$_{14}$ which is vastly insufficient to determine the 15 CF parameters. 
	
	Instead of fitting the CF parameters directly, we employ an effective PC model similar to the one demonstrated for pyrochlores.  Given the local two-fold symmetry, 9 independent parameters are required to fully describe the PC model for $R_{3}$Mg$_2$Sb$_{3}$O$_{14}$. These are $r_1$, $r_2$, $r_3$, $\theta_2$, $\theta_3$, $\phi_3$, $q_1$, $q_2$, $q_3$, with the detailed coordination of the eight surrounding PCs listed in Tab.~\ref{table:PC}. Compared to the 5 PC parameters for pyrochlores, the 4 additional parameters ($r_3$, $\theta_3$, $\phi_3$,  $q_3$) are associated with the O3 position that is split from O2 as a result of breaking the three-fold symmetry. Following the procedure established for pyrochlores, we  build an effective PC model using the crystallographic $R$-O-$R$ bond angles with $\theta_2\approx78^\circ$, $\theta_3\approx76.5^\circ$, $\phi_3\approx59^\circ$ \cite{Dun_2017}, and use the same reduction factor, $f\!\approx\!0.72$, for the PC distances, \ie $r_1 = r_2 = r_3 = f*r_1^c$. Compared to pyrochlores, the local environment for the oxygen atoms in the puckered-ring has changed dramatically. The O2  ion in a tripod-kagome structure is in the center of two $R^{3+}$, one Mg$^{2+}$, and one Sb$^{5+}$, whereas the O3 ion is in the center of one $R^{3+}$, one Mg$^{2+}$, and two Sb$^{5+}$. Since Sb$^{5+}$ captures a majority of the covalent electrons from $O^{2-}$, we expect a much smaller effective charge amount associated with O3 compared to O2. Thus, we choose $q_1 = 0.5e$, $q_2 = 0.3e$, $q_3 = 0.15e$ for the effective PC model of the tripod kagome structure.
	
	\subsection{PC fit to inelastic neutron scattering}
\begin{figure}[tbp]
		\linespread{1}
		\par
		\begin{center}
			\includegraphics[width= \columnwidth ]{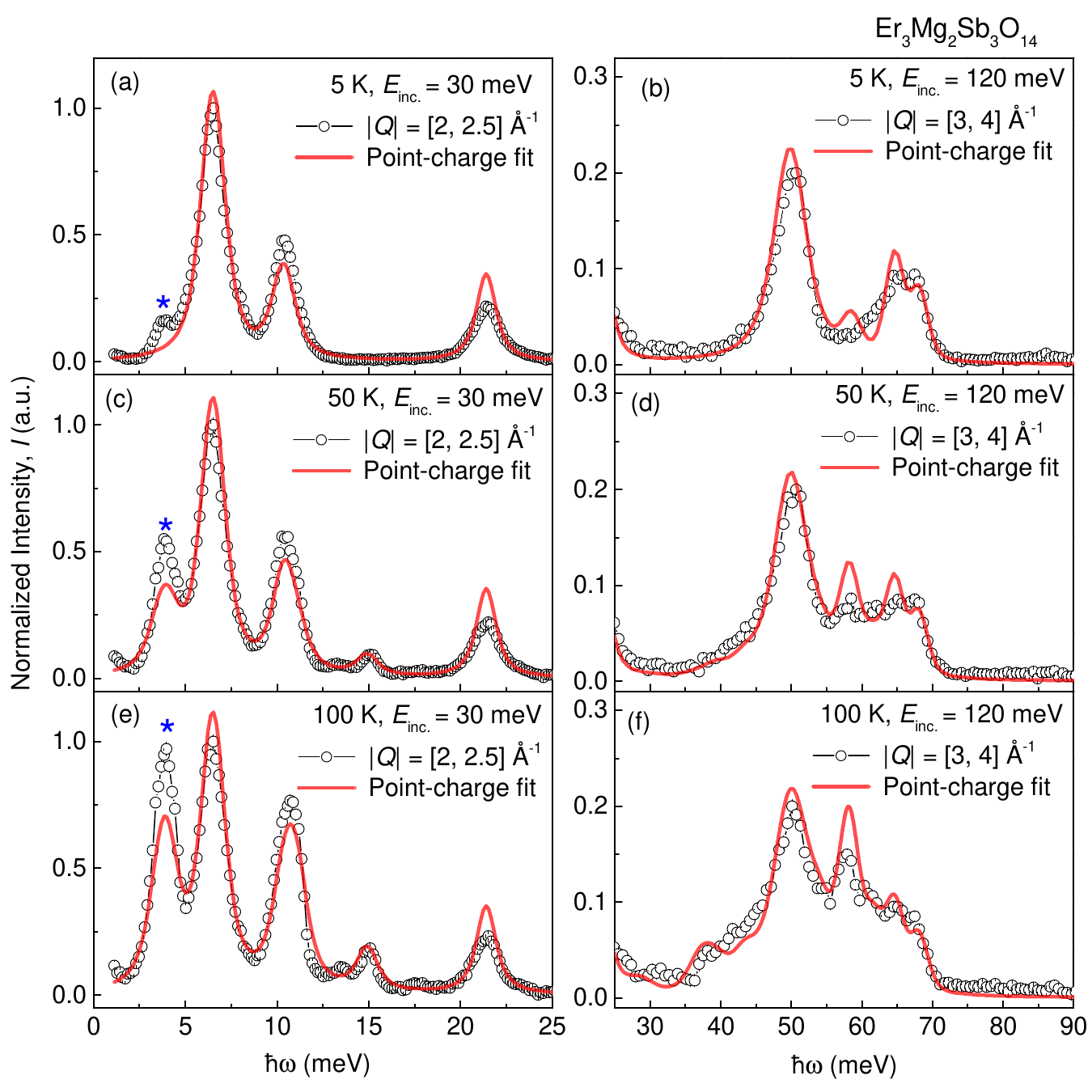}
		\end{center}
		\par
		\caption{\label{Fig:INS-Er}  Energy dependence of the measured CF excitations for Er$_{3}$Mg$_2$Sb$_{3}$O$_{14}$ (open black circles) and our best PC fits to data (solid red lines). The experimental data are extracted from the contour maps of Fig. \ref{Fig:INS} with the phonon background subtracted. The star labels an unexpected mode which is likely associated with a small percentage of the site-disordered Er at the Mg site.   }
	\end{figure}

\begin{figure}[tbp]
		\linespread{1}
		\par
		\begin{center}
			\includegraphics[width= \columnwidth ]{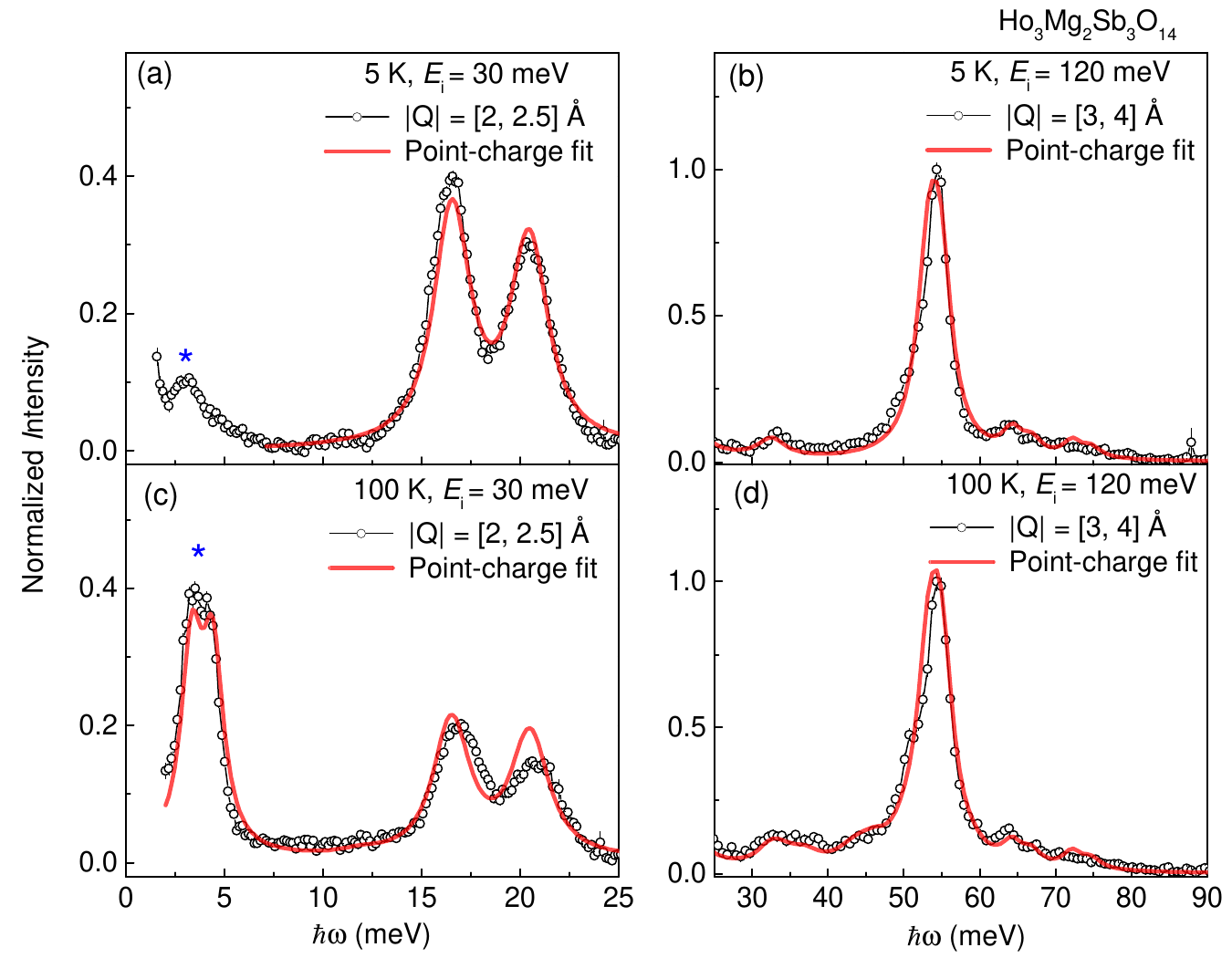}
		\end{center}
		\par
		\caption{\label{Fig:INS-Ho}  Energy dependence of the CF excitations in Ho$_{3}$Mg$_2$Sb$_{3}$O$_{14}$ (open black circles) and our best PC fits (solid red lines). The experimental data are extracted from the contour maps of Fig. \ref{Fig:INS} with the phonon background subtracted. The star labels an unexpected mode which is likely associated with a small percentage of the site-disordered Ho at the Mg site. }
	\end{figure}

\begin{figure}[tbp]
		\linespread{1}
		\par
		\begin{center}
			\includegraphics[width= 3 in ]{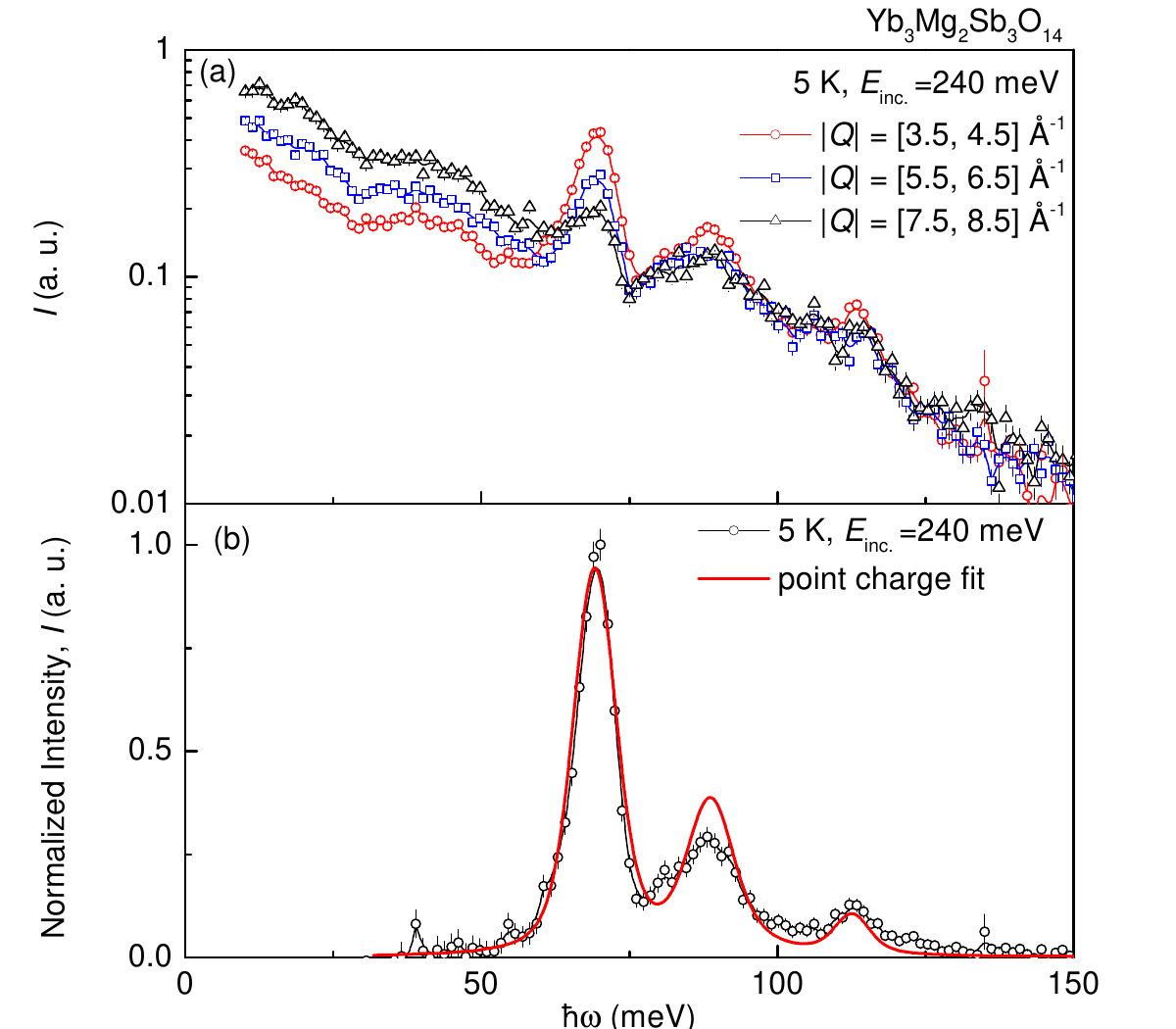}
		\end{center}
		\par
		\caption{\label{Fig:INS-Yb} (a) Energy dependence of the measured neutron spectra for Yb$_{3}$Mg$_2$Sb$_{3}$O$_{14}$ integrated in different momentum transfer windows. data are extracted from the contour maps of Fig. \ref{Fig:INS} and is plotted in a log-scale for a better illustration. (b) Phonon background subtracted data (open black circles) and our best PC fits (solid red line). }
	\end{figure}

\begin{figure}[tbp]
		\linespread{1}
		\par
		\begin{center}
			\includegraphics[width= \columnwidth ]{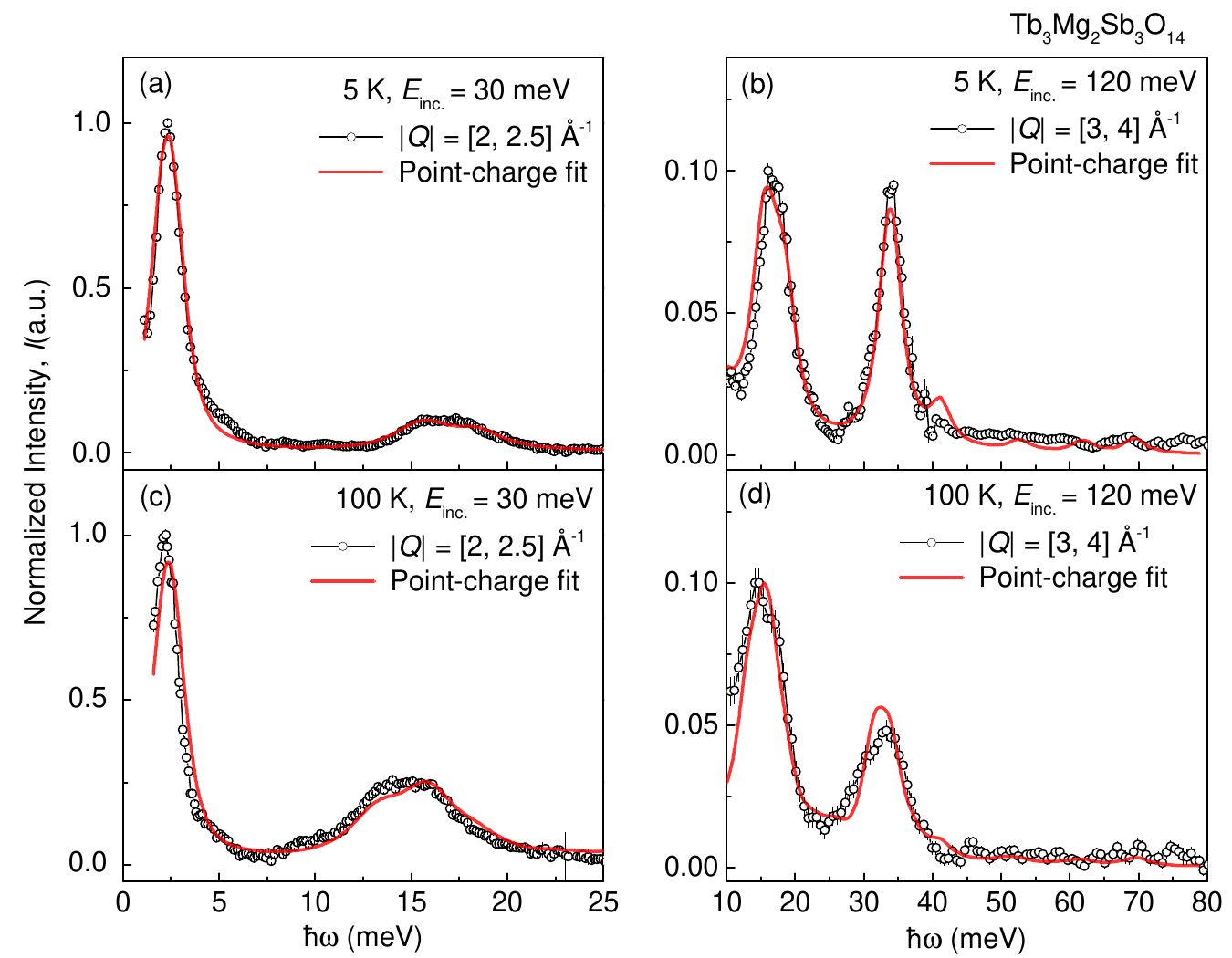}
		\end{center}
		\par
		\caption{\label{Fig:INS-Tb}  Energy dependence of the measured CF excitations for Tb$_{3}$Mg$_2$Sb$_{3}$O$_{14}$ (open black circles) and our best PC fits to data (solid red lines). The experimental data are extracted from the contour maps of Fig. \ref{Fig:INS} with the phonon background subtracted.}
	\end{figure}

	We refine the effective PC model by fitting inelastic neutron scattering spectra. An overview of the inelastic neutron scattering spectra for $R_3$Mg$_2$Sb$_3$O$_{14}$ ($R$ = Tb, Ho, Er, Yb) is shown in Fig. \ref{Fig:INS}. Four datasets are plotted for each compound, showing the excitation spectra measured at low ($T=5$\,K) and high ($T=100$\,K or $300$\,K) temperatures, and with several incident neutron energies. We clearly observe CF excitations from the ground-state, which intensities decay with $Q$, and mark them with blue arrows. Phonon excitations are only observed for Yb$_{3}$Mg$_2$Sb$_{3}$O$_{14}$ below 60 meV [Fig. \ref{Fig:INS}(d)] and subtracted according to Sec.~\ref{sec:methods}. Following this data analysis procedure, the energy-dependence of the signal is in Figs. \ref{Fig:INS-Er}--\ref{Fig:INS-Tb} for each compound, respectively.  
	
	We begin with Er$_{3}$Mg$_2$Sb$_{3}$O$_{14}$ ($J$ = 15/2)for which we expect seven CF excitation levels from the ground-state Kramers doublet. Four of the seven excitations are clearly seen below 55 meV at 5\,K at $\hbar\omega$ = 6.4(2), 10.5(3), 21.6(4), and 50(1) meV, respectively [Fig. \ref{Fig:INS} and Fig. \ref{Fig:INS-Er}(a-b)]. Between 55 meV and 70 meV, a somewhat continuous spectra is observed which is likely originating from three CF excitations in addition to some background. Within this energy range, two intensity maxima are observed at 65(1) and 67.5(9) meV. Instead of attempting to resolve the missing CF level, we perform a global fit to all the spectra in Fig. \ref{Fig:INS-Er} by vary the 9 PC parameters. The best fit to the data is achieved for $\theta_2 = 80.7^\circ$, $\theta_3 = 75.3^\circ$, $\phi = 59.9^\circ$, $q_1 = 0.511$, $q_2 = 0.311$, $q_3 = 0.187$, \blue{and a fitted Lorentzian peak width of $\gamma_L$ = 0.52 meV}, a solution that is not far from the initial model in parameter space. The fitted value of PC distances shows a trend opposite to the crystallographic $R$-O distances, with distances decreasing from $r_1$ =1.734 \AA\,to $r_2$ = 1.540 \AA, and $r_3$ =1.479 \AA. Given the large difference in atomic number between Mg$^{2+}$ and  Sb$^{5+}$ as well as the empty versus filled $d$-orbitals, this is not a surprise in light of the results on Er$_2X_2$O$_7$ discussed above. The CF parameters as well as the CF wave-functions for Er$_{3}$Mg$_2$Sb$_{3}$O$_{14}$ are listed in Tab.~\ref{table:Tripod_PC}. The curves fitted from the PC model agree well with all measured data-sets except for an intensity mismatch around 3.8 meV which shows up at both 5\,K and 100\,K (marked by stars). We can rule out magnetic impurities as well as the transition between the 6.4(2) to 10.5(3) meV. A similar weak peak is observed in Ho$_{3}$Mg$_2$Sb$_{3}$O$_{14}$ at $\sim$ 3 meV [Fig. \ref{Fig:INS-Ho}] , but not in Yb$_{3}$Mg$_2$Sb$_{3}$O$_{14}$ [Fig. \ref{Fig:INS}(d)], ruling out the possibility of an unsubtracted phonon signal. A likely explanation is structural site-disorder. Previous studies of the tripod kagome compounds generally indicate 3-5\% $R^{3+}$ ions located at the Mg$^{2+}$ site, which is surrounded by 6 oxygen ligands \cite{Dun_2017, Paddison_2016, Scheie2016effective, Scheie2018crystal} that likely gives rise to different CF excitations. A Lorentzian peak fit to the 5\,K, $E_\textrm{inc.} = 30\,meV$ spectra indicates a 3.4(3)\% peak intensity of the 3.4 meV feature compared to that of the strongest peak at 6.4(2) meV, consistent with the percentage of site-disorder. Furthermore, our effective PC model based on an octahedral ligand environment (Tab.~\ref{table:PC}) predicts that the most intense CF excitation is indeed around 3-6 meV for  site-disordered Er$^{3+}$, Tb$^{3+}$, and Ho$^{3+}$ ions, \blue{while the high intensity excitations expected for site-disordered Yb$^{3+}$ ions are around 60-70 meV \cite{PointChargeCF_2021}}. This scenario is further supported by recent CF measurements on spinel MgEr$_2$Se$_4$ where the strongest CF excitations is observed at 4.16 meV for Er in a similar ligand environment \cite{Plessis2019deviation}. 
	
	Next, we look at Ho$_{3}$Mg$_2$Sb$_{3}$O$_{14}$  ($J$ = 8) for which the CF spectra is expected to comprise 2$J$+1=17 singlet levels \cite{dun2020quantum}. However, instead of seeing 16 CF levels, our measurements resolve 5 crystal-field excitations [Fig. \ref{Fig:INS}(b), Fig. \ref{Fig:INS-Ho}], whose energy scheme and relative intensities resemble those of Ho$_2$Ti$_2$O$_7$ \cite{Rosenkranz_2000} except for an overall renormalization in energy. This is expected because as long as the deviation from trigonal symmetry is small, all the non-Kramers doublets in the pyrochlores should only split weakly in energy which is generally beyond the resolution of our neutron measurements. In this sense, it is almost impossible to perform a conventional CF fit based on the resolved CF energies and intensity ratios. Thus, we perform a global fit to the four spectra in  Fig. \ref{Fig:INS-Ho} to refine and effective PC model that best reflects the CF excitations. \blue{A grid search in parameter space near the fitted value for Er$_{3}$Mg$_2$Sb$_{3}$O$_{14}$ is first performed to find potential local minimums of $R_{wp}$}, which is used for choosing  initial values of our PC parameters. The best fit is shown as red lines in Fig. \ref{Fig:INS-Ho}.

	We continue with Yb$_{3}$Mg$_2$Sb$_{3}$O$_{14}$ for which Yb$^{3+}$ has $J$ = 7/2 for which we expect to see three CF excitations from the ground-state Kramers doublet. Our measurements indeed resolve three modes at 69.3(5) meV, 89(1), and 113(1) meV, respectively [Fig. \ref{Fig:INS}(d), Fig. \ref{Fig:INS-Yb}]. We isolate the pure CF signals by subtracting the low-$Q$ intensities with a fraction of the high-$Q$ intensities, and the normalized CF spectra is shown as black dots in Fig. \ref{Fig:INS-Yb} (b). Since the Boltzmann factor at 300\,K is not large enough to populate the higher CF levels, only the 5\,K data-set is used for the PC fit. In the current case, the number of experimental observable is considerably less than either the number of CF parameters (15) or that of PC parameters (9), so we expect a lot of degenerate solutions which would give us identical fits to the spectra. \blue{In an attempt to reduce the number of fit parameters, we fix the parameters associated with the angular distribution ($\theta_2, \theta_3, \phi$) and charge amount ($q_1, q_2, q_3$) to equal to the values defined by the effective point charge model.} By varying the PC parameters  $r_1, r_2, r_3$, the best fit to the experimental data is plotted as the red line in Fig. \ref{Fig:INS-Yb}(b), and the obtained PC and CF parameters are listed in Tab.~\ref{table:Tripod_PC}.

	Finally, we turn to Tb$_{3}$Mg$_2$Sb$_{3}$O$_{14}$.  The Tb$^{3+}$ ion has $J = 6$, for which we expect to see 13 singlet CF excitations at 5\,K. However, we observe only 3 groups of excitations [Fig. \ref{Fig:INS}(a)]. While the first excitation at 2.2(3) meV is relatively sharp, the latter two excitation modes are extremely broad and their main peak intensities are distributed between 10-20 meV and 30-40 meV, respectively. Similar to that of Yb$_{3}$Mg$_2$Sb$_{3}$O$_{14}$, we start from the PC parameters defined by the effective point charge model, and obtains a fit to data that fulfills the criteria $r_1 > r_2 > r_3$ and $ 0.5 \approx e_1 > e_2 > e_3 $.  Note that for this compound, the fitted value of $\gamma_L$ is considerably larger than that of the Er and Ho compounds. In addition, some over-fit to the background is noticeable above 40 meV [Fig. \ref{Fig:INS-Tb}] where the excitation spectra is dominated by phonons [Fig. \ref{Fig:INS}(a)]. 
	
	\begin{figure*}[tbp]
		\linespread{1}
		\par
		\begin{center}
			\includegraphics[width= 6.5 in ]{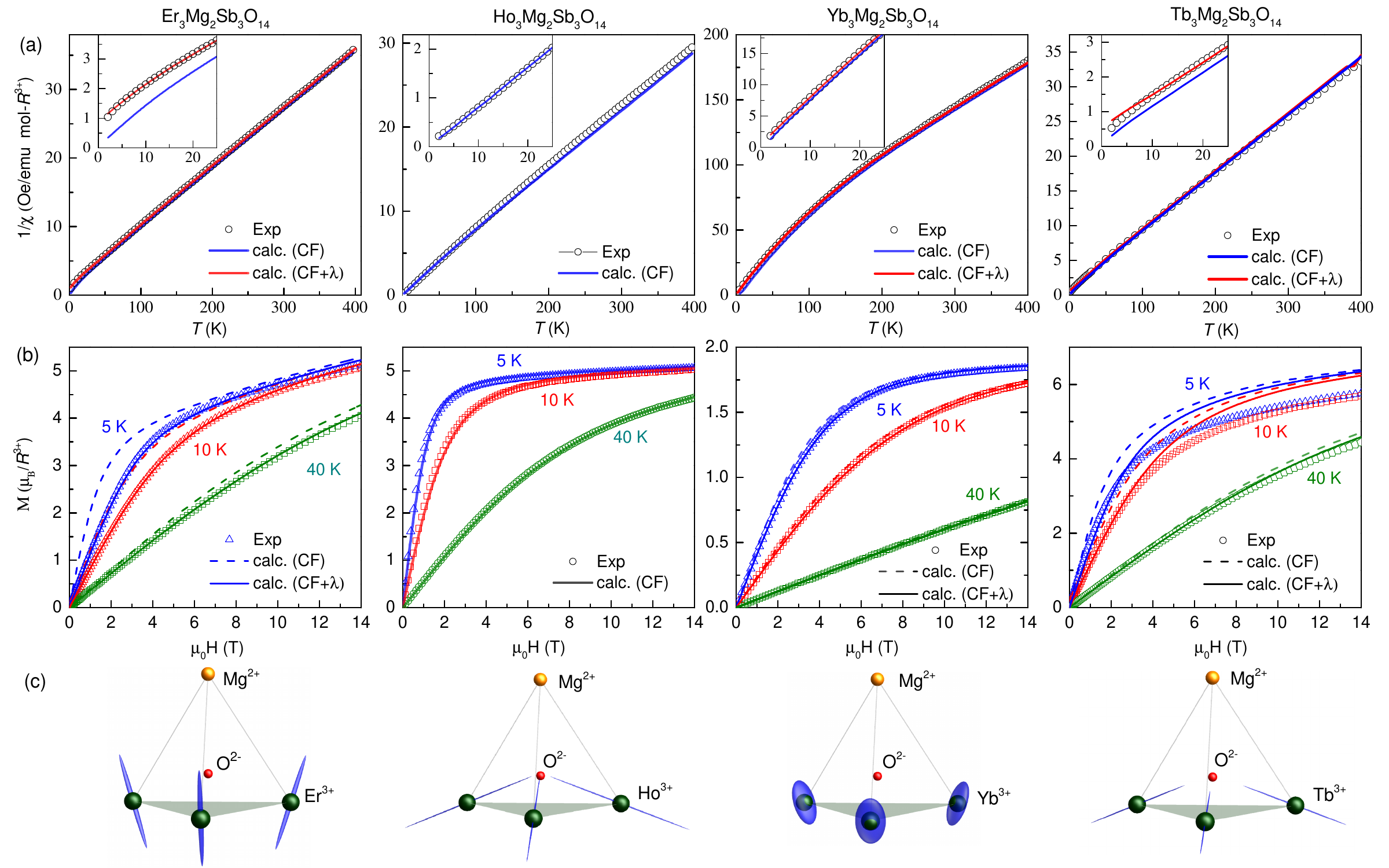}
		\end{center}
		\par
		\caption{\label{Fig:MH} (a) Measured (open circles) and calculated (solid lines) magnetic susceptibility for $R_{3}$Mg$_2$Sb$_{3}$O$_{14}$ ($R$ = Tb, Ho, Er, Yb). Red and blue lines denote the CF susceptibility with and without a molecular-field correction, respectively. (b)  Measured (open circles) and calculated (solid lines) isothermal magnetization curves at various temperatures. Solid and dashed lines denote the CF susceptibility with and without a molecular-field correction, respectively. (c) Orientations and geometries of the local $g$-tensor ellipsoids with respect to the kagome plane.}
	\end{figure*}
	
\subsection{ Susceptibility/magnetization \& $g$-tensor/principle-axes}
Table \ref{table:Tripod_PC} lists the fitted values of the PC parameters, the corresponding CF parameters, and CF wave-functions for $R_{3}$Mg$_2$Sb$_{3}$O$_{14}$ ($R$ = Tb, Ho, Er, Yb).  Using these parameters, we calculate the powder-averaged DC susceptibility ($\chi\rm{^{CF}_{powder}}$) and isothermal magnetization ($M\rm{^{CF}_{powder}}$) from the CF levels of each of the compounds. As shown in Fig. \ref{Fig:MH} (a), the calculated $1/\chi\rm{^{CF}_{powder}}$ and $M\rm{^{CF}_{powder}}$ generally agrees with the measured curves for all temperatures and magnetic fields which strongly validates our approach. When taking a closer look at the data for Er$_{3}$Mg$_2$Sb$_{3}$O$_{14}$,  $1/\chi\rm{^{CF}_{powder}}$ seems to underestimate the measured values by a constant amount [Fig. \ref{Fig:MH}(a) inset], meanwhile, $M\rm{^{CF}_{powder}}$ tends to overestimate the measured magnetization, which becomes more obvious at low temperatures [Fig. \ref{Fig:MH}(b)].  Both disagreements can be explained by two-ion antiferromagnetic interactions. Once we account for this effect using a Weiss molecular field, the corrected susceptibility ($\chi\rm{^{CF+\lambda}_{powder}}$) and  magnetization ($M\rm{^{CF+\lambda}_{powder}}$) with $\lambda$ = 0.321\,K almost perfectly agrees with the experimental values. The same correction can be made for the other three compounds.  We obtain the value of $\lambda$  by fitting the susceptibility below 25\,K which yields -0.012\,K for Ho$_{3}$Mg$_2$Sb$_{3}$O$_{14}$, 0.302\,K for Yb$_{3}$Mg$_2$Sb$_{3}$O$_{14}$, and  0.215\,K for Tb$_{3}$Mg$_2$Sb$_{3}$O$_{14}$. We notice that for the Tb compound, the corrected $1/\chi\rm{^{CF+\lambda}_{powder}}$ still clearly deviates from the experimental curve below 10 K. Meanwhile, although $M\rm{^{CF+\lambda}_{powder}}$ accounts for the magnetization at 40\,K quite well, it obviously overestimates magnetization at high-field and low-temperature. Therefore, we conclude that while our PC fits successfully describes the CF Hamiltonian for  $R_{3}$Mg$_2$Sb$_{3}$O$_{14}$ ($R$ =  Ho, Er, Yb), further investigations are necessary to determine unambiguously the CF Hamiltonian for Tb$_{3}$Mg$_2$Sb$_{3}$O$_{14}$.  \blue{It is also possible that the simple assumption of molecular field breaks down, which is likely associated with the low-lying CF and interactions between the multipolar degrees of freedom (as will discussed below). } 
	
\begin{figure}[tbp]
\linespread{1}
\par
\begin{center}
\includegraphics[width= \columnwidth ]{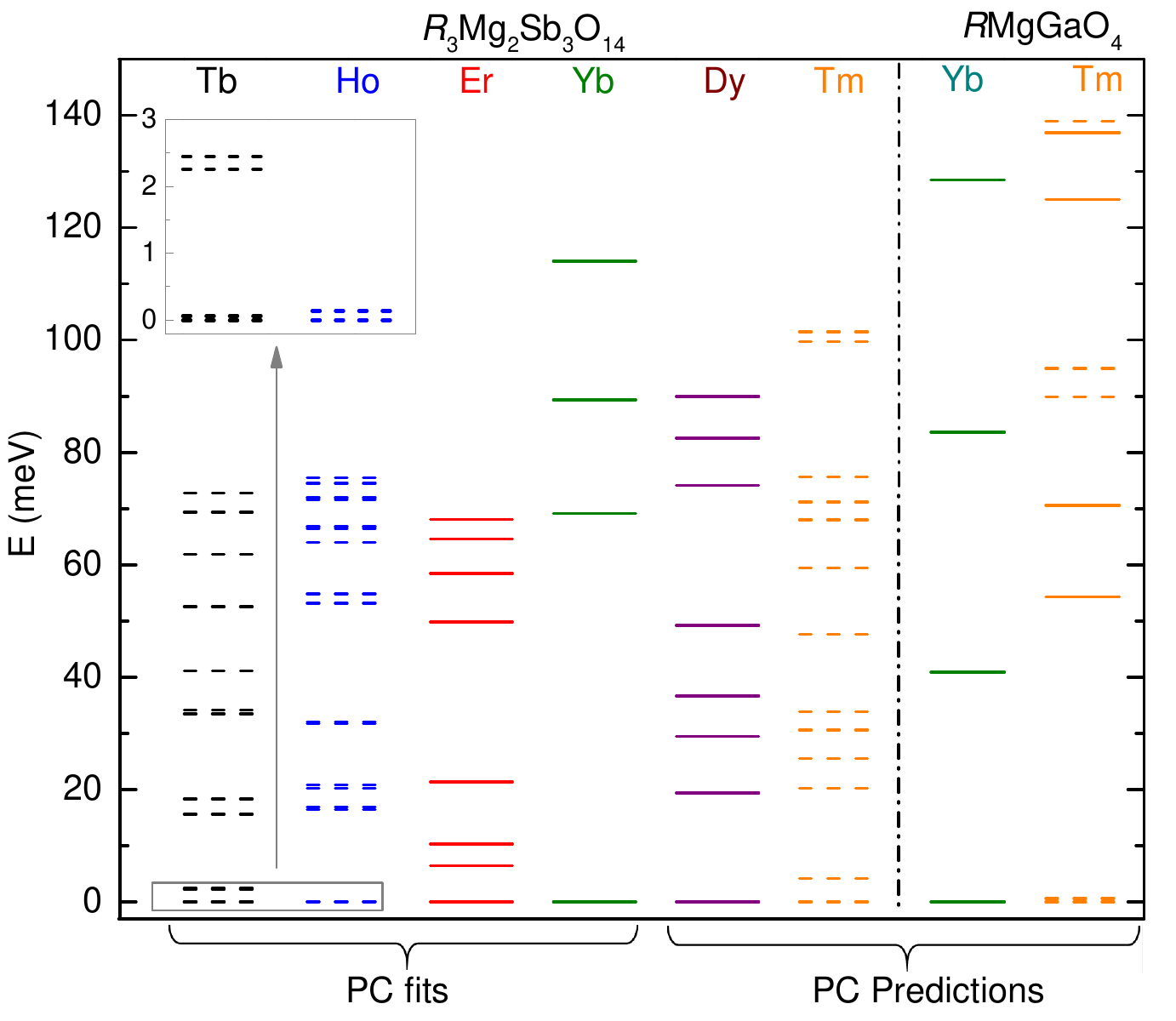}
\end{center}
\par
\caption{\label{Fig:Predicton} A summary of the fitted and predicted CF levels from the effective PC model for six tripod kagome compounds $R_{3}$Mg$_2$Sb$_{3}$O$_{14}$, and two rare earth triangular compounds $R$MgGaO$_{4}$.  Solid and dashed lines present doublet and singlet levels, respectively. Inset: zoomed low energy region below 3 meV for Tb$_{3}$Mg$_2$Sb$_{3}$O$_{14}$ and Ho$_{3}$Mg$_2$Sb$_{3}$O$_{14}$. }
	\end{figure}
	
	The CF scheme for the four compounds from our PC fit is summarized in Fig. \ref{Fig:Predicton}. For Er and Yb compounds, the CF ground state is a well isolated Kramers doublet (designated by $\ket{\pm}$). For Tb and Ho compounds, the CF ground-state comprises two singlets (designated by $\ket{0}$ and $\ket{1}$) that are weakly split in energy. Importantly, $\ket{0}$ and $\ket{1}$ can be approximately expressed in the symmetric and anti-symmetric form of a non-Kramers doublet,
	\begin{eqnarray}
	\ket{0} \approx \frac{1}{\sqrt{2}} (\ket{+}+\ket{-}), \;\;
	\ket{1} \approx \frac{1}{\sqrt{2}} (\ket{+}-\ket{-}). \label{eq:singlets}
	\end{eqnarray} 
	Since symmetric and anti-symmetric wave-functions are the eigenstates of the  $\sigma^x$ Pauli matrix,  an energy splitting between two crystal-field singlets can be exactly mapped into a transverse magnetic field acting on a corresponding doublet \cite{Wang_1968, dun2020quantum}. Therefore, the mapping from the total angular momentum basis to the effective spin-1/2 basis is still valid as long as the two-singlet (or quasi-doublet) is well separated from higher-energy CF levels. Eq.~\ref{eq:singlets} allows us to obtain the necessary wave-functions $\ket{\pm}$ that can be used in Eq. \ref{eq:pseudospin}-\ref{eq:g-factor} to calculate the $g$-tensor. As discussed earlier, with under-determined principal axes, a two-step rotation is required to make the $g$-tensor diagonal. Taking Er$_{3}$Mg$_2$Sb$_{3}$O$_{14}$ as an example, the calculated and the diagonalized $g$-tensor yield:
	\begin{eqnarray} \label{eq:g-factor-Ertripod}
	\small{
		\bm{g} = \left(\begin{array}{ccc} 13.55 & 0 & -3.54 \\ 0 & -0.178 & 0 \\ 0 & 0 &  0.50 \end{array}\right)
		\xrightarrow{}
		\left(\begin{array}{ccc} 14.05 & 0 & 0 \\ 0 & 0.18 & 0 \\ 0 & 0 &  0.50
		\end{array}\right), \nonumber
	}
	\end{eqnarray} 
	which are related by a pseudo-spin rotation (Eq. \ref{eq:rotationA}) and a pseudo/real-spin co-rotation (Eq.\ref{eq:rotationB}). The latter contains a rotation of 4.6$^\circ$ about the $C_2$-axis  which finally transforms the  $xyz$ coordination defined for our PC model into the principal coordinate $x'y'z'$ where $g$ is diagonal [Fig. \ref{Fig:Structure}(b)]. The real-space rotation angles required for the other three compounds can be obtained in the same way which are generally within 5$^\circ$ [see Table \ref{table:Tripod_PC}]. This suggests that our initial assumption is roughly correct: the O1 ligand provides the strongest Coulomb potential that distinguishes them from the remaining oxygen ligands, making the $R$-O1 bond direction approximately one of the principal axes. 
	
	\begin{table*}[tbp]  
		\setlength{\tabcolsep}{0.4em} 
		\renewcommand{\arraystretch}{1.2}
		\centering
		\caption{\label{table:Tripod_PC}  Tabulated results of the PC parameters, diagonalized $g$-tensors, CF parameters, CF energies, and wave-functions of the CF ground states for $R_{3}$Mg$_2$Sb$_{3}$O$_{14}$. The results for compounds with $R$ = Tb, Ho, Er, Yb are obtained from PC fits of Fig. \ref{Fig:INS-Er} to \ref{Fig:INS-Yb}. The numbers for Dy$_{3}$Mg$_2$Sb$_{3}$O$_{14}$ and Tm$_{3}$Mg$_2$Sb$_{3}$O$_{14}$ are predicted results from Effective PC models.   }
		\begin{tabular}{c|c|cccccccccc|cccc}
			\hline\hline 
			\multirow{2}{*}{ Compound} & \multirow{2}{*}{Method} & 	\multicolumn{10}{c|}{ PC parameters}   & \multicolumn{4}{c}{$g$-tensor}  \\
			& & $r_1$(\AA) & $r_2$(\AA) & $r_3$(\AA) & $\theta_2 (^\circ)$ & $\theta_3 (^\circ)$& $\phi_3 (^\circ)$ &$q_1 (e)$ & $q_2 (e)$  & $q_3 (e)$ &   $\gamma_L$ (meV) & $g_{xx}$ & $g_{yy}$ & $g_{zz}$ & $\gamma$($^\circ)$ \\
			\hline 
			Tb$_{3}$Mg$_2$Sb$_{3}$O$_{14}$ & \multirow{4}{*}{PC Fit} & 1.735 & 1.579 & 1.351 & 83.2 & 77.5 & 55.4 & 0.54 &  0.26 & 0.10  & 1.65 &  0 & 0 & 14.25&  1.0 \\
			Ho$_{3}$Mg$_2$Sb$_{3}$O$_{14}$  & & 1.639 & 1.572 & 1.395 & 81.1 & 79.6 & 54.7 & 0.504 &  0.302 & 0.122 & 1.05 & 0 & 0 & 19.46 & 0.2 \\
			Er$_{3}$Mg$_2$Sb$_{3}$O$_{14}$  & & 1.741 & 1.550 & 1.481 & 80.7 & 75.3 & 59.9 & 0.511 &  0.311 &  0.187 & 0.52 & 14.05  & 0.18 & 0.50 & 4.6  \\
			Yb$_{3}$Mg$_2$Sb$_{3}$O$_{14}$  & & 1.542  & 1.560 & 1.484 & 78.0 $^a$  & 76.5 $^a$  & 59 $^a$  & 0.5$^a$  &  0.3 $^a$  & 0.15 $^a$  & 1.50 & 5.21  & 3.42 & 1.58 & 1.6 \\
			\hline
			Dy$_{3}$Mg$_2$Sb$_{3}$O$_{14}$ &  \multirow{2}{*}{PC Calc.} & 1.68 & 1.57 & 1.38  & 78 & 76.5 & 59 & 0.5 & 0.3 & 0.15 & - & 0 & 0 & 19.13 & 4.5 \\
			Tm$_{3}$Mg$_2$Sb$_{3}$O$_{14}$ & & 1.64 & 1.55 & 1.48  & 78 & 76.5 & 59 & 0.5 & 0.3 & 0.15 & - & 0 & 0 & 0 &  -  \\
			\hline  	
			\multicolumn{16}{l}{\footnotesize{$^a$ This number is fixed during fitting}} \\
			\hline 
		\end{tabular}

		 \smallskip
		
		\renewcommand{\arraystretch}{1.1}
		\setlength{\tabcolsep}{0.35em} 
		\begin{tabular}{c|c|cccccccccccccccc}
			\hline
			\centering
			  Compound &  Method   & \multicolumn{15}{c}{CF parameters (meV)} \\
			  &  & $A_2^0$ & $A_2^1$ & $A_2^2$ & $A_4^0$ & $A_4^1$ & $A_4^2$ & $A_4^3$ & $A_4^4$ & $A_6^0$ & $A_6^1$ & $A_6^2$ & $A_6^3$ & $A_6^4$ & $A_6^5$ & $A_6^6$ \\
			\hline
			Tb$_{3}$Mg$_2$Sb$_{3}$O$_{14}$ &  \multirow{4}{*}{PC fit} &  39.70 & -6.21 & -24.48 & 34.53 & -5.39 & 4.98 & 221.36 & -52.08 & 2.90 & 6.32 & -4.91 & -79.50 &  15.09 & 80.30 & 97.63\\
			Ho$_{3}$Mg$_2$Sb$_{3}$O$_{14}$ & &  33.09 & -50.47 & -16.00 & 34.94 & 10.37 & -0.43 & 243.96 & -54.56 & 3.94 & -1.47 & -0.32 & -77.65 &  9.38 & 57.78 & 85.09 \\
			Er$_{3}$Mg$_2$Sb$_{3}$O$_{14}$ & &  -4.81 & 6.51 & -57.37 & 27.39 & -1.91 & 17.19 & 290.52 & -22.24 & 2.21 & -0.66 & -8.94 & -83.42 & 9.17 &  12.34 & 90.74 \\
			Yb$_{3}$Mg$_2$Sb$_{3}$O$_{14}$ & &  48.00 & -36.31 & -40.82 & 33.15 & 13.03 & 8.92 & 262.00 & -27.96 & 6.37 & -4.60 & -2.19 & -67.12 &  4.23 & -4.37 & 64.20\\
			\hline
			Dy$_{3}$Mg$_2$Sb$_{3}$O$_{14}$ &  \multirow{2}{*}{PC Calc.} &  31.79 & -50.89 &  -57.14 & 28.23 & 14.82 & 10.04 & 285.17 & -31.34 & 4.29 & -5.61 & -2.55 & -76.42 &  4.96 & -6.46 & 73.12\\
			Tm$_{3}$Mg$_2$Sb$_{3}$O$_{14}$ &  &  22.85 & -31.90 & -38.08 & 32.31 & -0.12 & 3.10 & 371.63 & -13.46 & 4.86 & 5.16 & -2.11 & -115.24 &  2.04 & 43.73 & 109.18\\
			\hline
	\end{tabular}
		 
		 \smallskip
		 
		\begin{tabular}{c|c|ll}
			\hline
			\centering
			& & & $\qquad$ CF eigen-energies (meV) and ground state eigen-functions in $\ket{J, J_z}$ basis\\
			\hline
			\multirow{5}{*}{Tb$_{3}$Mg$_2$Sb$_{3}$O$_{14}$ } &  \multirow{5}{*}{PC fit} 
			& $E_i =$ &  \; 0,\; 0.06,\; 2.3,\; 2.5,\; 15.6,\; 18.4, \; 33.5, \; 34.2,\; 41.1, \; 52.6, \; 61.9, \; 69.4, \;72.8 \\
			& & $\ket{0} =$ & $ 0.032(\ket{6}-\ket{-6})+0.682(\ket{5}+\ket{-5})- 0.012(\ket{4}-\ket{-4})$ \\
			& & & $- 0.137(\ket{3}+\ket{-3})-0.103(\ket{2}-\ket{-2})+0.049(\ket{1}+\ket{-1})-0.000\ket{0}$\\ 
			& & $\ket{1} =$ & $ 0.039(\ket{6}+ \ket{-6})+0.669(\ket{5}-\ket{-5})-0.133(\ket{4}+\ket{-4})$\\
			& & & $- 0.121(\ket{3}-\ket{-3})-0.125(\ket{2}+\ket{-2}+0.049(\ket{1}+\ket{-1})+0.027\ket{0}$ \\
			\hline
			\multirow{5}{*}{Ho$_{3}$Mg$_2$Sb$_{3}$O$_{14}$} &  \multirow{5}{*}{PC fit} 
			& $E_i =$ &  \; 0, 0.14, 16.4, 16.9, 20.2, 20.8, 31.7, 3208,  53.2,  54.8,  64.0,  66.4,  66.8, 71.5,  71.9, 74.6, 75.5 \\
			& & $\ket{0} = $ & $0.690(\ket{8}+\ket{-8})-0.004(\ket{7}-\ket{-7})-0.005(\ket{6}+\ket{-6})-0.120(\ket{5}-\ket{-5})$ \\
			& & & $+ 0.051(\ket{4}+\ket{-4})+0.020(\ket{3}-\ket{-3})+0.003(\ket{2}+\ket{-2})+0.069(\ket{1}-\ket{-1})-0.023\ket{0}$ \\
			& & $\ket{1} = $ & $0.693(\ket{8}-\ket{-8})-0.016(\ket{7}+\ket{-7})+0.002(\ket{6}-\ket{-6})-0.107(\ket{5}+\ket{-5})$\\
			& & & $ -0.010(\ket{4}-\ket{-4})-0.067(\ket{3}+\ket{-3})+0.031(\ket{2}-\ket{-2})-0.049(\ket{1}-\ket{-1})-0.000\ket{0}$ \\
			\hline
			\multirow{5}{*}{Er$_{3}$Mg$_2$Sb$_{3}$O$_{14}$} & \multirow{5}{*}{PC fit} 
			& $E_i =$&  \; 0,\; 6.5,\; 10.3,\; 21.4,\; 49.8,\; 58.5, \; 64.6, \; 68.1 \\
			& & $\ket{\pm} = $ & $\pm0.074\ket{\pm15/2} -0.172\ket{\pm13/2}\pm0.217\ket{\pm11/2}+ 0.188\ket{9/2} \mp 0.252\ket{\pm7/2} $ \\
			& & & $ +0.325\ket{\pm5/2} \mp0.309\ket{\pm3/2}+0.265\ket{\pm1/2}\mp0.404\ket{\mp1/2} + 0.095\ket{\mp3/2} \mp  0.050\ket{\mp5/2}$ \\
			& & & $   -0.383\ket{\mp7/2}\pm0.317\ket{\mp9/2}-0.240\ket{\mp11/2}\pm0.253\ket{\mp13/2} -0.061\ket{\mp15/2}$ \\
			\hline
			\multirow{3}{*}{Yb$_{3}$Mg$_2$Sb$_{3}$O$_{14}$}  & \multirow{3}{*}{PC fit} 
			&  $E_i =$ & \;0,\; 69.1,\; 88.3, \; 115.4 \\
			& & $\ket{\pm} = $ & $ 0.037\ket{\pm7/2} \mp 0.082\ket{\pm5/2} -0.105\ket{\pm3/2} \pm 0.021\ket{\pm1/2} -0.947\ket{\mp1/2} $ \\
			& & & $  \pm 0.050\ket{\mp3/2} -0.087\ket{\mp5/2} \pm0.272\ket{\mp7/2}$ \\  
			\hline
			\multirow{4}{*}{Dy$_{3}$Mg$_2$Sb$_{3}$O$_{14}$} & \multirow{4}{*}{PC Calc.} 
			& $E_i =$  & \; 0,\;19.4,\; 29.5, \; 36.7\; 49.3, \; 74.2, \; 82.6, \; 90.0 \\
			& & $\ket{\pm} = $ & $ \pm0.034\ket{\pm15/2}+0.015\ket{\pm13/2}\mp0.001\ket{\pm11/2}- 0.010\ket{9/2} \pm 0.025\ket{\pm7/2}$ \\
			& & & $   -0.010\ket{\pm5/2} \mp0.088\ket{\pm3/2}-0.002\ket{\pm1/2}\pm0.051\ket{\mp1/2} + 0.006\ket{\mp3/2} \mp 0.059\ket{\mp5/2}$ \\
			& & & $   +0.047\ket{\mp7/2}\pm0.259\ket{\mp9/2}+0.038\ket{\mp11/2}\mp0.126\ket{\mp13/2} -0.948\ket{\mp15/2}$ \\   
			\hline
			\multirow{3}{*}{Tm$_{3}$Mg$_2$Sb$_{3}$O$_{14}$} & \multirow{3}{*}{PC Calc.} 
			& $E_i =$  & \; 0,\; 4.2,\; 20.3,\; 25.5,\;30.6,\; 33.9, \; 47.6, \; 59.4, \; 68.0, \; 71.2, \; 75.7 \; 99.8, \; 101.5 \\
			& & $\ket{0} =  $ & $-0.116(\ket{6}-\ket{-6})+0.105(\ket{5}-\ket{-5})+ 0.097(\ket{4}-\ket{-4})$ \\
			& & & $   +0.649(\ket{3}+\ket{-3})+0.034(\ket{2}-\ket{-2}) + 0.208(\ket{1}+\ket{-1})-0.000\ket{0}$\\ 
			\hline\hline  	
	\end{tabular}
	\end{table*}
	
	\subsection {Single-ion \& Collective physics}
	In this section, we provide a case by case discussion of the the implications of our work for the collective physics in each of the four tripod-kagome compounds.
	
	The CF two-singlet in Ho$_{3}$Mg$_2$Sb$_{3}$O$_{14}$ gives rise to an effective Ising moment with Ising axes pointing into the center of local tetrahedron [Fig. \ref{Fig:MH}(c)]. The strong dipolar interactions in such an arrangement prefer one-in-two-out or two-in-one-out configurations in a triangle, which gives rise to emergent magnetic charges and classical spin-fragmentation physics at low temperatures \cite{Chern_2011, Paddison_2016}. Along with the effective transverse field generated from the splitting of 0.15 meV of the two-singlet CF ground state, the systems maps into a canonical model for quantum magnetism: interacting Ising spins under a transverse field.  As demonstrated in our separate study \cite{dun2020quantum}, a transverse Ising model based on a dipolar kagome ice is promising to stabilize a high-entangled quantum state, and  Ho$_{3}$Mg$_2$Sb$_{3}$O$_{14}$ provides an example on how quantum fluctuations can be generated from CF effects alone. 
	
	Similarly, Tb$^{3+}$ ions in Tb$_{3}$Mg$_2$Sb$_{3}$O$_{14}$ possess a two-singlet CF ground state with a separation of 0.06 meV, according to our PC fits. Although this number might be not be accurate, the anisotropy of pseudo-spins is expected to be the same Ising type as that of Ho$_{3}$Mg$_2$Sb$_{3}$O$_{14}$ if restricted to the lowest two singlets [Fig. \ref{Fig:MH}(c)]. Interestingly, unlike Ho$_{3}$Mg$_2$Sb$_{3}$O$_{14}$, neither a magnetic order nor emergent magnetic charge order in observed experimentally in Tb$_{3}$Mg$_2$Sb$_{3}$O$_{14}$ \cite{Dun_2017}. The difference is likely coming from the two low-lying excited CF singlets at 2.0 meV and 2.6 meV, which is absent in Ho$_{3}$Mg$_2$Sb$_{3}$O$_{14}$. 
As these levels are comparable in energy to the spin-spin interactions, a proper effective description of the CF will necessarily include 4 singlet levels, akin to the virtual CF excitation theory proposed for Tb$_2$Ti$_{2}$O$_{7}$ \cite{molavian2007dynamically}.  
We emphasize that determining the CF Hamilton for Tb-related compounds has been proven to be a challenging task. Even within the pyrochlore family, the CF Hamiltonian for Tb-pyrochlores has received the most attention and controversy over the last decade \cite{Mirebeau2007magnetic, Zhang2014neutron, princep2015crystal}. This is due, first, to weak CF levels that are not easily determined by neutron experiments which complicates the CF fits; second, the high energy CF excitations are strongly contaminated by phonons where magneto-elastic coupling is likely to play a key role. Similar magneto-elastic coupling have been demonstrated to be related to the peak splitting of the strongest CF excitations in Ho-pyrochlores \cite{gaudet2018magnetoelastically}. For Tb$_{3}$Mg$_2$Sb$_{3}$O$_{14}$,  we notice that the high-energy phonon signals are much more intense (in absolute scale) than that of the other three compounds, providing an anomalous background that prevents us determining any CF levels above 70 meV [Fig. \ref{Fig:INS}(a-d)]. 
	
	A previous study has shown that Yb$_{3}$Mg$_2$Sb$_{3}$O$_{14}$ exhibits long ranged magnetic order at 1.67\,K  with a magnetic structure that has not been reported to date~\cite{Dun_2017}.  With $g_{xx} > g_{yy}  > g_{zz}$, the anisotropy ellipsoid of Yb$^{3+}$ has an almond shape where none of the three components in $g$-tensor is negligible [Fig. \ref{Fig:MH} (c)]. Since the super-exchange can be obtained as a perturbation to the CF Hamiltonian, we expect that the exchange interactions that couple to the $xy$ components of spins will be stronger than others associated with the $z$ component, with possible existence of off-diagonal couplings between $xy$ and $z$ components. Therefore, the magnetic structure is unlikely to be the all-in-all-out type as that observed in Nd$_{3}$Mg$_2$Sb$_{3}$O$_{14}$ \cite{Scheie2016effective}, but rather likely to be a non-coplanar magnetic structure with spins showing a great tendency to lie in the local $xy$ plane. 
	
	The most surprising result of our work is that the $g$-tensor of Er$_{3}$Mg$_2$Sb$_{3}$O$_{14}$ is very uniaxial. This contradicts our previous assumptions of XY anisotropy which was based on a comparison to Er-pyrochlore analogues \cite{Dun_2016}. However, this Ising-like anisotropy has two fundamental differences compared to that of the Ho or Tb compounds: first, the easy axis is not pointing along the local $z$-direction, but rather $x$-direction that is about 65$^\circ$ canted away from the kagome plane; second, the components of $g_{yy}$ and $g_{zz}$ are not zero which in principle allow for observable spin dynamics even in the absence of transverse fields. The two consequences are, first, the system is still highly frustrated given an antiferromagnetic interaction between the moments along local $x$-direction; second, similar to that in the Yb and Er pyrochlores \cite{Ross_2011, Savary_2012}, we expect off-diagonal couplings between $x$ components and $yz$ components. Since no magnetic ordering down to 50 mK has been observed experimentally \cite{Dun_2016, Dun_2017}, Er$_{3}$Mg$_2$Sb$_{3}$O$_{14}$ is an quantum spin liquid candidate whose interpretation calls for a quantum Ising model with anisotropic exchanges. 
	
	\begin{table*}[tbp]  
		\setlength{\tabcolsep}{0.5em} 
		\renewcommand{\arraystretch}{1.2}
		\centering
		\caption{\label{table:YMGO} Tabulated CF parameters, CF energies, and $g$-tensors from the Effective PC calculations for the average structure of YbMgGaO$_4$ and  TmMgGaO$_4$. }
		\begin{tabular}{c|ccc|cccccc|c|cc}
			\hline\hline 
			PC Calc. &	\multicolumn{3}{c|}{ PC parameters}  & \multicolumn{6}{|c|}{CF parameters (meV)$^a$ } & CF energies & \multicolumn{2}{c}{$g$-tensor}  \\
			& $r_1$(\AA)  & $\theta_1 (^\circ)$ & $q_1 (e)$  & $A_2^0$ & $A_4^0$ & $A_4^3$ & $A_6^0$ & $A_6^3$ & $A_6^6$ & $E_i$  (meV) & $g_\perp$ & $g_\parallel$  \\
			\hline
			YbMgGaO$_4$  & 1.607  & 61.7 & 0.5  & -58.1 & -15.1 & 695.0 & 5.1 & -34.6 & 52.0 & 0, 41.3, 84.3, 129.6 & 3.32 & 2.10\\
			\multirow{2}{*}{TmMgGaO$_4$} & \multirow{2}{*}{1.613}  & \multirow{2}{*}{61.7} & \multirow{2}{*}{0.5}  &
			\multirow{2}{*}{-77.5} & \multirow{2}{*}{-15.7} & \multirow{2}{*}{721.7} & \multirow{2}{*}{5.5} & \multirow{2}{*}{-36.9} & \multirow{2}{*}{55.4} & Singlet: 0, 0.4, 87.0, 92.1, 134.4 & \multirow{2}{*}{0} & \multirow{2}{*}{13.24} \\
			& & & & & & & & & & Doublet: 52.5, 68.5, 120.6, 132.3 & \\
			\hline\hline  	
			\multicolumn{6}{l}{\footnotesize{$^a$Here, $B_n^m =  A_n^m\theta_n$ according to Eq. \ref{eq:Steven}.}}
		\end{tabular}
	\end{table*}
	
	\section {Discussion: other applications}
	\begin{figure}[tbp]
		\linespread{1}
		\par
		\begin{center}
			\includegraphics[width= \columnwidth ]{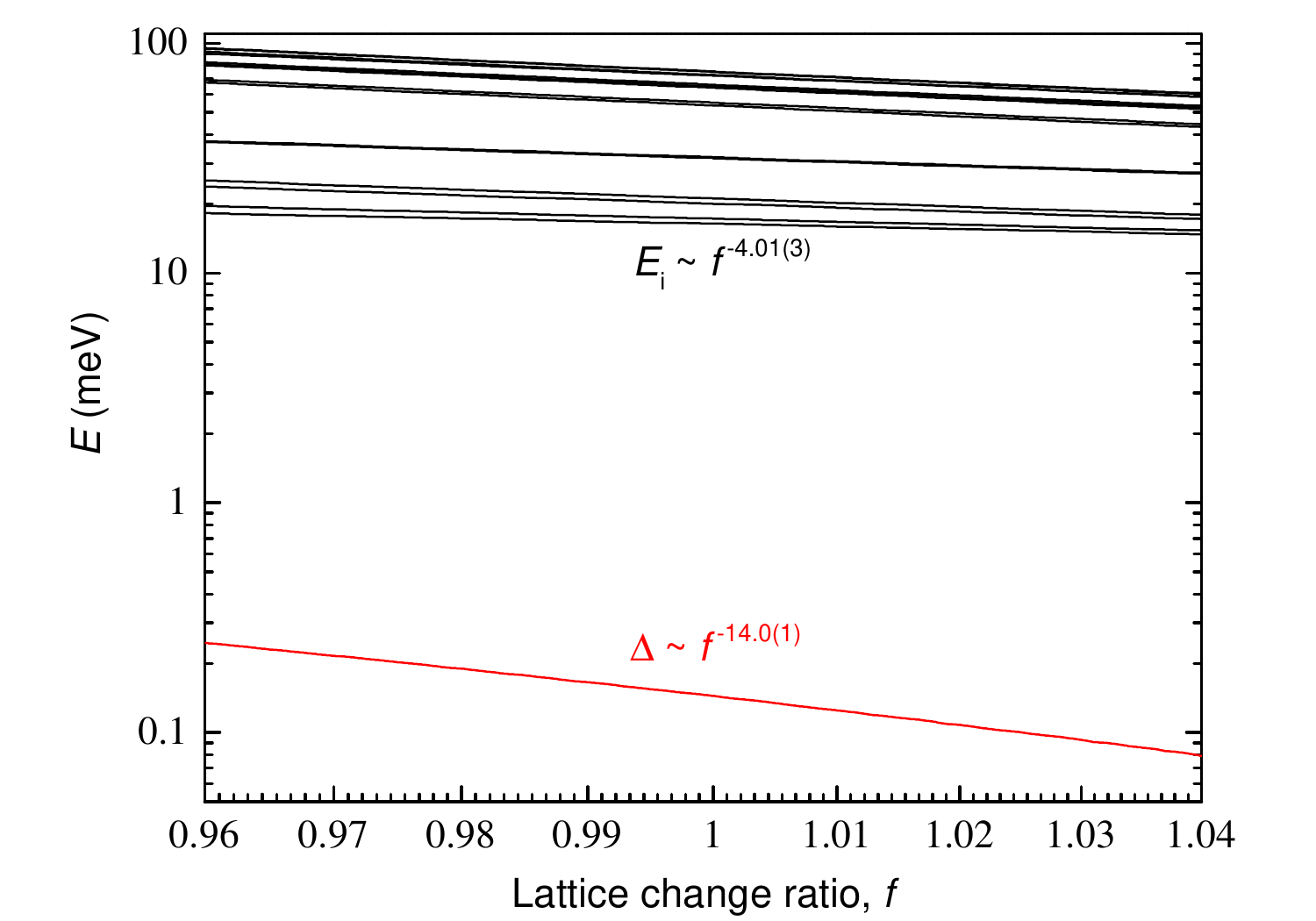}
		\end{center}
		\par
		\caption{\label{Fig:Pressure}  CF energy of Ho$_{3}$Mg$_2$Sb$_{3}$O$_{14}$  as a function of lattice change in a log-log scale.  The lattice contraction/expansion is characterized by a ratio, $f$, where $f$ =1 corresponds to the fitted PC model in Table \ref{table:Tripod_PC}. By performing a linear fit of the two-singlet splitting between $f$ = 0.96 and 1.04, we obtained a exponential dependence as  $\Delta \sim f^{-14.0(1)}$.  }
	\end{figure}
	
	\subsection{Tuning the transverse field in Ho$_{3}$Mg$_2$Sb$_{3}$O$_{14}$ }
	Similar to the example established for Yb$_2$Ti$_2$O$_7$ in Fig. \ref{Fig:Dependence}, the effective PC model can be used to study how the CF reacts to the changes of PC parameters. This can sometimes be extremely useful. Taking Ho$_{3}$Mg$_2$Sb$_{3}$O$_{14}$ for example, our previous work has shown that the excitation associated with the splitting of the ground-state doublet in Ho$_{3}$Mg$_2$Sb$_{3}$O$_{14}$ appears highly overdamped due to interactions between sites, which precludes a direct measurement of that energy scale with neutrons \cite{dun2020quantum}. To minimize the interaction effects between sites and to directly observe the two-singlet excitation, we synthesized a very dilute Ho tripod kagome compound (Ho$_{0.01}$La$_{0.99}$)${_3}$Mg${_2}$Sb${_3}$O${_{14}}$ whose  single-ion excitation can be understand within a two singlet splitting of $\Delta$=1.14 \,K \cite{dun2020quantum}. As the lattice parameters the doped compound are approximately 2.9\% larger than that of Ho${_3}$Mg${_2}$Sb${_3}$O${_{14}}$, a natural question that arises is how $\Delta$ responds to the change in lattice parameters. This can be easily investigated by our PC calculations. If we assume the  PC parameters for distance ($r_i$) changes linearly with the lattice contraction/expansion, we can track the CF energies as a function of a lattice change. By varying the lattice parameters nearby the fitted PC parameters for Ho$_{3}$Mg$_2$Sb$_{3}$O$_{14}$, denoted by a ratio, $f$, all the excited CF levels have an exponential dependence of the lattice contraction/expansion, evidenced by the linear dependence of the CF energies as a function of $f$ in a log-log plot [Fig. \ref{Fig:Pressure}].  Notably, the two singlet splitting has a much more dramatic response to lattice changes compared to other excited CF levels, \ie $\Delta \sim f^{-14.0(1)}$ versus $E_{i} \sim f^{-4.01(3)}$. It means that if we extrapolate the value of $\Delta$ from (Ho$_{0.01}$La$_{0.99}$)${_3}$Mg${_2}$Sb${_3}$O${_{14}}$ to Ho${_3}$Mg${_2}$Sb${_3}$O${_{14}}$, the corresponding value of $\Delta$ will be 1.14$\times$(1.029)$^{14}$ = 1.72\,K.  It also means that one can efficiently tune the transverse field in Ho$_{3}$Mg$_2$Sb$_{3}$O$_{14}$ by applying physical or chemical pressure: a 3\% change in lattice will result in 50\% change in $\Delta$. Recalling the $1/f^3$ dependence of the dipolar interaction, it means lattice contraction/expansion can effectively tune the ratio of transverse field over the dipolar interactions in Ho$_{3}$Mg$_2$Sb$_{3}$O$_{14}$. Importantly, this ratio is the essence of the transverse Ising model on a kagome dipolar magnet which determines the boundary between several distinct quantum phases, as demonstrated by our previous simulations \cite{dun2020quantum}. This example shows how PC calculations can be used to investigate the effects of physical or chemical pressure on the quantum dynamics of transverse Ising model, which also provides insights into the search for quantum spin ices based on non-Kramers ions. 
	
	\subsection{Scaling CF to Dy$_{3}$Mg$_2$Sb$_{3}$O$_{14}$ and Tm$_{3}$Mg$_2$Sb$_{3}$O$_{14}$}
	In the spirit of the Effective PC model, we can make predictions of CF for other tripod kagome compounds. This is similar to the scaling arguments proposed for the pyrochlores but within a more physically-meaningful framework \cite{bertin2012crystal}.  Starting from the Effective PC model for the tripod structure, we can choose $q_1 = 0.5e$, $q_2 = 0.3e$, $q_3 = 0.15e$, and $\theta_2 = 78^\circ$, $\theta_3 = 76.5^\circ$, $\phi_3 = 59^\circ$ which are the approximate numbers from the local crystallography. For the distance PC parameters, we extrapolate the fitted $r_i$ values from $R_{3}$Mg$_2$Sb$_{3}$O$_{14}$ ($R$ = Tb, Ho, Er, Yb)  since Dy and Tm are between Tb/Ho and Er/Yb  in the periodic table, respectively. The numbers for the Effective PC model are listed in the Table \ref{table:Tripod_PC} and the predicted CF levels are plotted in  Fig. \ref{Fig:Predicton}. Consistent with earlier studies, the Dy$^{3+}$ ion exhibits a well isolated Kramers doublet ground state where the first excited CF level is around 32 meV \cite{Paddison_2016}. On the other hand, Tm$_{3}$Mg$_2$Sb$_{3}$O$_{14}$ is expected to host a well isolated singlet ground state with an energy separation of 6 meV from the first excited singlet. This energy separation is expected to be at least one order of magnitude larger than the spin-spin interactions, making interesting many-body physics irrelevant at low temperature. Therefore, similar to that of Pr$_{3}$Mg$_2$Sb$_{3}$O$_{14}$ \cite{Dun_2017, Scheie2018crystal}, we expect only single-ion magnetism for Tm$_{3}$Mg$_2$Sb$_{3}$O$_{14}$. We comment that a tripod kagome variant, Tm$_{3}$Zn$_2$Sb$_{3}$O$_{14}$, whose low temperature magnetism has been investigated recently \cite{ding2018possible}, is likely to be associated with both a large two CF singlets splitting and site-disorder effects which are enhanced for Zn-based tripod kagome compounds to due the large $R$ atomic number \cite{Dun_2017}.
	
	\subsection{Triangular magnets TmMgGaO$_4$ and YbMgGaO$_4$}
	Given the success of our effective PC model to treat the pyrochlore and tripod kagome compounds, it is natural to extend the model for other rare earth oxides, for example, YbMgGaO$_4$ and TmMgGaO$_4$. While it is known that there is some structure disorders associated with Mg/Ga occupying the same site \cite{Li_2015_PRL, Paddison_2017, li2017crystalline}, both compounds share the same average structure where the rare-earth ions sit in an octahedral environment with $D_{3d}$ point group, and oxide ligands are in the center of a tetrahedron formed by three rare earth ion and one non-magnetic ion. Although it requires the same number of Stevens' parameters to describe its CF as for the pyrochlores, many fewer PC parameters are needed to describe the local ligand geometry (see Table \ref{table:PC}).  Similar to the Effective PC model established in the previous sections, we chose the three PC parameters, $r_1 = 0.72\,r_1^c$, $\theta_1$ = 61.7$^{\circ}$, and $q_1 = 0.5\,e$. The predicted CF levels for YbMgGaO$_4$ and TmMgGaO$_4$ are tabulated in Table \ref{table:YMGO} and plotted in Fig. \ref{Fig:Predicton}. Interestingly, TmMgGaO$_4$ exhibits the same two-singlet ground state as Ho$_{3}$Mg$_2$Sb$_{3}$O$_{14}$ and Tb$_{3}$Mg$_2$Sb$_{3}$O$_{14}$, where the 0.4 meV splitting of the two-singlet can be mapped to a transverse field that is comparable to spin-spin interactions. The estimated splitting is close to the experimental number based on the heat capacity measurement of a diluted sample \cite{li2018absence} and also agrees with the number from theoretical fits to the spin dynamics \cite{shen2019intertwined, li2020kosterlitz}, suggesting that the Effective PC model is doing a good job at capturing the CF levels of the average structure. In contrast, the predicted CF levels for YbMgGaO$_4$ shows three excited levels at 40.9, 83.6, and 128.5 meV, respectively. While the former two levels are close in energy to the 38 and 97 meV levels measured experimentally \cite{Paddison_2017, li2017crystalline}, the other CF level (at 128.5 meV) is much higher in energy than that observed experimentally (at 61 meV). Interestingly, our previous measurements indeed observe an additional weak CF signal at 134 meV~\cite{Paddison_2017}, and this observation is also reported by Li \textit{et al.} but interpreted as a possible neutron multiple scattering effect \cite{li2017crystalline}. The discrepancies between PC model and experiments indicate that the CF Hamiltonian of YbMgGaO$_4$ needs to be revisited with a different approach.  \\

	\section {Summary \& Outlook }
	In summary, this manuscript introduces the concept of effective PC model to understand the CF excitations measured by inelastic neutron scattering in rare-earth based magnets. It provides a clear methodology that is benchmarked successfully to rare-earth pyrochlore oxides and subsequently applied to tripod kagome magnets. We predict that our approach will be useful in practice for systems with low point-group symmetry where a large number of CF parameters are expected. Compared to the conventional Stevens' operator approach widely used in the last few decades, the advantages of the PC approach includes a physically meaningful parameterization, a reduced parameter space, and a level of accuracy sufficient for CF excitations to be utilized to determine the principal axes of the spin-space anisotropy tensor. Perhaps the most exciting message from our results is that a reasonable estimation of the CF spectra and spin-space anisotropies can be made simply from the local crystallography provided PC parameters are adjusted. Our manuscript provides a methodology to achieve that adjustment. Given the current emphasis on magnetic quantum matter with anisotropic spins and exchange interactions, we expect our PC calculation method to be useful in searching for rare-earth materials with desired anisotropy properties. 
	
	\begin{acknowledgments}
		The authors thank J.~A.~M Paddison for many helpful discussions and Huibo Cao for critical reading of the manuscript. The work of Z.L.D., X.B. and M.M. at Georgia Tech was supported by the U.S. Department of Energy, Office of Science, Office of Basic Energy Sciences, Neutron Scattering Program under Award Number DE-SC0018660. The work of H.D.Z. at the University of Tennessee was supported by the U.S. Department of Energy, Office of Science, Office of Basic Energy Sciences, Materials Sciences and Engineering Division under award DE-SC-0020254. The research at Oak Ridge National Laboratory's Spallation Neutron Source was sponsored by the U.S. Department of Energy, Office of Basic Energy Sciences, Scientific User Facilities Division.
	\end{acknowledgments}

	\bibliography{PointCharge.bib}

\end{document}